\documentclass{article}
\pdfoutput=1
\usepackage{arxiv}

\usepackage{amssymb}
\usepackage{latexsym}

\usepackage{url}
\usepackage{xcolor}

\usepackage{titling}
\predate{}
\postdate{}

\usepackage[numbers]{natbib}
\usepackage{hyperref}

\usepackage{booktabs}
\usepackage{tabu}

\usepackage{amsmath}

\usepackage{graphicx}
\usepackage{subcaption}
\usepackage{placeins}

\usepackage{soul}

\usepackage{wrapfig}
\usepackage{authblk}

\title{A Multi-site Study of a Breast Density Deep Learning Model for Full-field Digital Mammography Images and Synthetic Mammography Images}

\author[1]{Thomas P. Matthews}
\author[1]{Sadanand Singh}
\author[1]{Brent Mombourquette}
\author[1]{Jason Su}
\author[1]{Meet P. Shah}
\author[1]{Stefano Pedemonte}
\author[1]{Aaron Long}
\author[2]{David Maffitt}
\author[2]{Jenny Gurney}
\author[1]{Rodrigo Morales Hoil}
\author[1]{Nikita Ghare}
\author[1]{Douglas Smith}
\author[2]{Stephen~M.~Moore}
\author[3]{Susan C. Marks}
\author[2]{Richard L. Wahl}

\affil[1]{Whiterabbit AI, Inc., Santa Clara, CA 95054, USA}
\affil[2]{Mallinckrodt Institute of Radiology, Washington University School of Medicine,  St. Louis, MO 63110, USA }
\affil[3]{Peninsula Diagnostic Imaging, San Mateo, CA 94401, USA}

\date{}

\begin{document}

\begin{titlepage}
    \vspace*{2cm}
    \begin{center}
    This manuscript has been accepted for publication in Radiology: Artificial Intelligence (https://pubs.rsna.org/journal/ai), which is published by the Radiological Society of North America (RSNA).
    
    (\copyright~2020 RSNA)
    \end{center}
\end{titlepage}

\newpage

\maketitle
\setstcolor{red}
\setul{}{0.5ex}

\begin{abstract}

\textbf{Purpose}: To develop a Breast Imaging Reporting and Data System (BI-RADS\textsuperscript{\textregistered}) breast density deep learning (DL) model in a multi-site setting for synthetic two-dimensional mammography (SM) images derived from digital breast tomosynthesis exams using full-field digital mammography (FFDM) images and limited SM data. \\

\textbf{Materials and Methods}: A DL model was trained to predict BI-RADS breast density using FFDM images acquired from 2008 to 2017 (Site~1: 57492 patients, 187627 exams, 750752 images) for this retrospective study. The FFDM model was evaluated using SM datasets from two institutions (Site~1: 3842 patients, 3866 exams, 14472 images, acquired from 2016 to 2017; Site~2: 7557 patients, 16283 exams, 63973 images, 2015 to 2019). Each of the three datasets were then split into training, validation, and test datasets. Adaptation methods were investigated to improve performance on the SM datasets and the effect of dataset size on each adaptation method is considered. Statistical significance was assessed using confidence intervals (CI), estimated by bootstrapping. \\

\textbf{Results}: Without adaptation, the model demonstrated substantial agreement with the original reporting radiologists for all three datasets (Site~1 FFDM: linearly-weighted $\kappa_w$~=~0.75 [95\% CI: 0.74, 0.76]; Site~1 SM: $\kappa_w$~=~0.71 [95\% CI: 0.64, 0.78]; Site~2 SM: $\kappa_w$~=~0.72 [95\% CI: 0.70, 0.75]). With adaptation, performance improved for Site~2 (Site~1: $\kappa_w$~=~0.72 [95\% CI: 0.66, 0.79], 0.71 vs 0.72, P~=~.80; Site~2: $\kappa_w$~=~0.79 [95\% CI: 0.76, 0.81], 0.72 vs 0.79, P~$<$~.001) using only 500 SM images from that site. \\

\textbf{Conclusion}: A BI-RADS breast density DL model demonstrated strong performance on FFDM and SM images from two institutions without training on SM images and improved using few SM images. \\

\textbf{Key Results}:
\begin{itemize}
    \item A Breast Imaging Reporting and Data System breast density deep learning (DL) model achieved substantial agreement with the original interpreting radiologists for full-field digital mammography (FFDM) exams from Site 1 (linearly-weighted Cohen’s kappa: $\kappa_w$~=~0.75 [95\% confidence interval (CI): 0.74, 0.76]). \\
    \item Without modification, the DL model trained on FFDM images demonstrated substantial agreement with the original reporting radiologists for a test set of synthetic two-dimensional mammography (SM) images, which are generated as part of digital breast tomosynthesis exams (Site~1: $\kappa_w$~=~0.71 [95\% CI: 0.64, 0.78]). \\
    \item Without modification, the DL model also demonstrated close agreement for a test set of SM images obtained from a different institution than that of the training data (Site~2: $\kappa_w$~=~0.72 [95\% CI: 0.70, 0.75]). Adaptation techniques requiring few SM images were able to further improve performance (eg Site~2: $\kappa_w$~=~0.79 [95\% CI: 0.76, 0.81], P~$<$~.001).
\end{itemize}
\vspace{1mm}

\textbf{Summary}: A breast density deep learning model showed strong performance on digital and synthetic mammography images from two institutions without training on synthetic mammography images and improved with adaptation using few synthetic mammography images. \\

\textbf{Abbreviations:} BI-RADS~=~Breast Imaging Reporting and Data System, CI~=~ confidence interval, DBT~=~digital breast tomosynthesis, DL~=~deep learning, FFDM~=~full-field digital mammography, SM~=~synthetic two-dimensional mammography, 2D~=~two-dimensional, 3D~=~three-dimensional
\end{abstract}

\section{Introduction} \label{sec:introduction}

Breast density is an important risk factor for breast cancer \cite{kerlikowske_breast_2010, mccormack_breast_2006, boyd_mammographic_2007}. Additionally, areas of higher density can mask findings within mammograms leading to lower sensitivity \cite{mandelson_breast_2000}. Many states have passed breast density notification laws requiring clinics to inform women of their breast density \cite{kressin_content_2016}. Radiologists typically assess breast density using the Breast Imaging Reporting and Data System (BI-RADS) lexicon, which divides breast density into four categories: A, almost entirely fatty; B, scattered areas of fibroglandular density; C, heterogeneously dense; and D, extremely dense (examples are presented in Figure~\ref{fig:example_density_cview}) \cite{sickles_acr_2013}. Unfortunately, radiologists exhibit intra- and inter-reader variability in the assessment of BI-RADS breast density, which can result in differences in clinical care and estimated risk \cite{sprague_variation_2016, spayne_reproducibility_2012, berg_breast_2000}. \\

Deep learning (DL) has previously been employed to assess BI-RADS breast density for film \cite{yi_deep-learning_2018} and full-field digital mammography (FFDM) images \cite{gandomkar_bi-rads_2019, mohamed_deep_2018, ma_multi-path_2019, lehman_mammographic_2019, wu_breast_2017, youk_automated_2016} with some models demonstrating closer agreement with consensus
estimates than individual radiologists \cite{lehman_mammographic_2019}. To realize the promise of using these DL models in clinical practice, two key challenges must be met. First, as breast cancer screening is increasingly moving to digital breast tomosynthesis (DBT) \cite{richman_adoption_2019} due to improved reader performance \cite{friedewald_breast_2014, skaane_comparison_2013, rafferty_diagnostic_2014}, DL models should be compatible with DBT exams. To aid in radiologist interpretation of breast cancer and breast density, DBT exams have two-dimensional (2D) images in addition to three-dimensional (3D) images. These 2D images may be either FFDM images or synthetic 2D mammography (SM) images derived from the 3D images. Figure~\ref{fig:cview_ffdm_comparison} shows the differences in image characteristics between FFDM and SM images. The relatively recent adoption of DBT at many institutions means that the datasets available for training DL models are often fairly limited for DBT exams compared with FFDM exams. Second, DL models must offer consistent performance across sites, where differences in imaging technology, patient demographics, or assessment practices could impact model performance. To be practical, this should be achieved while requiring little additional data from each site. \\

In this study, we present a BI-RADS breast density DL model that offers close agreement with the original reporting radiologists for both FFDM and DBT exams at two institutions. A DL model was first trained to predict BI-RADS breast density using a large-scale FFDM dataset from one institution. Then, the model was evaluated on a test set of FFDM exams as well as SM images generated as part of DBT exams acquired from the same institution and from a separate institution. Adaptation techniques, requiring few SM images, were explored to improve performance on the two SM datasets.

\section{Materials and Methods} \label{sec:materials_and_methods}

This retrospective study was approved by an institutional review board for each of the two sites where data were collected (Site~1: internal institutional review board and Site~2: Western Institutional Review Board, Puyallup, WA).
Informed consent was waived, and all data were handled according to the Health Insurance Portability and Accountability Act. This work was supported in part by funding from Whiterabbit AI, Inc. Washington University has equity interests in Whiterabbit AI, Inc. and may receive royalty income and milestone payments from a “Collaboration and License Agreement” with Whiterabbit AI, Inc. to develop a technology evaluated in this research.

\subsection{Datasets} \label{sec:datasets} 

Mammography exams were collected from two sites: Site~1, an academic medical center located in the midwestern region of the United States, and Site~2, an out-patient radiology clinic located in northern California. For Site~1, separate FFDM and SM datasets were collected, while for Site~2, only a SM dataset was collected. The Site~1 FFDM dataset consisted of 187627 exams acquired from 2008~to~2017, the Site~1 SM dataset consisted of 3866 exams acquired from 2016 to 2017, and the Site~2 SM dataset consisted of 16283 exams acquired from 2015 to 2019. The FFDM images were acquired on Hologic Selenia and Selenia Dimensions imaging systems (Hologic,~Inc., Marlborough, MA), while the SM image were from Hologic Selenia Dimension imaging systems (C-View, Hologic,~Inc., Marlborough, MA). The two sites serve different patient populations. The patient cohort from Site~1 consisted of 59\%~White (34192~of~58397), 23\%~African~American (13201~of~58397), 3\%~Asian (1630~of~58397), and 1\%~Hispanic (757~of~58397) while Site~2 is 58\%~White (4350~of~7557), 1\%~African~American (110~of~7557), 21\%~Asian (1594~of~7557), and 7\%~Hispanic (522~of~7557). The distribution of ages is similar for the two sites (Site~1, 55~$\pm$~16~years; Site~2, 56~$\pm$~11~years). 

The exams were interpreted by one of 11 radiologists with breast imaging experience ranging from 2 to 30 years for Site~1 and by one of nine radiologists with experience ranging from 10 to 41 years for Site~2. The BI-RADS breast density assessments of the radiologists were obtained from each site's mammography reporting software (Site~1, Magview 7.1, Magview, Burtonsville, Maryland; Site~2, MRS 7.2.0; MRS Systems Inc. Seattle, Washington). Patients were randomly selected for training, validation, and testing at ratios of 80\%, 10\%, and 10\%, respectively. Since the split was performed at the patient-level, the images for a given patient appear in only one of these sets. All exams with a BI-RADS breast density assessment were included. No explicit filtering was performed for implants or prior surgery. For the FFDM validation set, only the first 25000 images were used in order to accelerate the training process, as evaluation on the validation set occurs after each training epoch. For the test sets, exams were required to have exactly the four standard screening mammography images (the mediolateral oblique and craniocaudal views of both breasts). This restriction led to the elimination of nearly all exams with implants due to the presence of implant displaced views. Following these restrictions, the distribution of patients was the following: training (FFDM, 50700 [88\%]; Site~1 SM, 3169 [82\%]; Site~2 SM, 6056 [80\%]), validation (FFDM, 1832 [3\%]; Site~1 SM, 403 [10\%]; Site~2 SM, 757 [10\%]), and testing (FFDM, 4960 [9\%]; Site~1 SM, 270 [7\%]; Site~2 SM, 744 [10\%]). The distribution of the BI-RADS breast density assessments for each set are presented in Table~\ref{tab:wustl_datasets} (Site~1) and Table~\ref{tab:pdi_datasets} (Site~2).

\begin{table}[htbp]
    \centering
    \caption{Description of the Site~1 FFDM and SM Training, Validation, and Test Datasets}
    \begin{tabular}{lllllll}
        \vspace*{1mm}\\
        \toprule
         ~ & \multicolumn{3}{c}{FFDM} & \multicolumn{2}{c}{SM} \\
         \toprule
         \cmidrule(l){2-4} \cmidrule(l){5-6} 
         ~ & Training & Validation & Test & Training & Validation & Test \\
         \midrule
         Patients, n & 50700 (88\%) & 1832 (3\%) & 4960 (9\%) & 3169 (82\%) & 403 (10\%) & 270 (7\%) \\
         Exams, n & 168208 & 6157 & 13262 & 3189 & 407 & 270 \\
         Images, n & 672704 & 25000 & 53048 & 11873 & 1519 & 1080 \\
         \midrule
         \multicolumn{2}{l}{BI-RADS Category} & & & & & \\
         \midrule
         ~~A &  80459 ~(12.0\%) & 3465 (13.9\%) & 4948 (9.3\%) & 1160 (9.8\%) & 154 (10.1\%) & 96 (8.9\%) \\
         ~~B & 348878 (51.9\%) & 12925 (51.7\%) & 27608 (52.0\%) & 6121 (51.6\%) & 771 (50.8\%) & 536 (49.6\%) \\
         ~~C & 214465 (31.9\%) & 7587 (30.3\%) & 18360 (34.6\%) & 3901 (32.9\%) & 510 (33.6\%) & 388 (35.9\%) \\
         ~~D & 28902 ~(4.3\%) & 1023 (4.1\%) & 2132 (4.0\%) & 691 (5.8\%) & 84 (5.5\%) & 60 (5.6\%) \\
         \bottomrule
    \end{tabular}
    \vspace*{3mm}
    \caption*{Note.-- BI-RADS category distribution for images is shown on bottom. BI-RADS = Breast Imaging-Reporting Data System, FFDM = full-field digital mammography, SM = synthetic two-dimensional mammography}
    \label{tab:wustl_datasets}
\end{table}

\begin{table}[htbp]
    \centering
    \caption{Description of the Site~2 SM Training, Validation, and Test Datasets}
    \begin{tabular}{llll}
        \vspace*{1mm}\\
        \toprule
         ~ & Training & Validation & Test  \\
         \midrule
         Patients, n & 6056 (80\%) & 757 (10\%) & 744 (10\%)  \\
         Exams, n & 13061 & 1674 & 1548 \\
         Images, n & 51241 & 6540 & 6192 \\
         \midrule
         BI-RADS Category & & & \\ 
         \midrule
         ~~A & 7866 (15.4\%) & 865 (13.2\%) & 948 (15.3\%) \\
         ~~B & 20731 (40.5\%) & 2719 (41.6\%) & 2612 (42.2\%) \\
         ~~C & 15706 (30.7\%) & 2139 (32.7\%) & 1868 (30.2\%) \\
         ~~D & 6938 (13.5\%) & 817 (12.5\%)& 764 (12.3\%) \\
         \bottomrule
    \end{tabular}
    \vspace*{3mm}
    \caption*{Note.-- BI-RADS category distribution for images is shown on bottom. BI-RADS = Breast Imaging and Reporting Data System, SM = synthetic two-dimensional mammography}
    \label{tab:pdi_datasets}
\end{table}

\subsection{Deep Learning Model} \label{sec:deep_learning_model} 
The DL model and training procedure were implemented using the pytorch DL framework (pytorch.org, version~1.0). The base model architecture is a pre-activation Resnet-34 \cite{he_identity_2016, he_deep_2016, wu_group_2018}, which accepts as input a single image corresponding to one of the views from a mammography exam and produces estimated probabilities that the image belongs to each of the BI-RADS breast density categories. The model was trained using the FFDM dataset following the procedure described in Appendix~\ref{sec:training_procedure}.

\subsection{Domain Adaptation Methods} \label{sec:domain_adaptation} 

The goal of domain adaptation is to take a model trained on a dataset from one domain (source domain) and transfer its knowledge to a dataset in another domain (target domain), which is typically much smaller in size. Features learned by DL models in the early layers can be general, ie domain and task agnostic \cite{yosinski_how_2014}. Depending on the similarity of domains and tasks, even deeper features learned from one domain can be reused for another domain or task. \\

In this work, we explore approaches for adapting the DL model trained on FFDM images (source domain) to SM images (target domain) that reuse all the features learned from the FFDM domain. First, inspired by the work of Guo~et~al~\cite{guo_calibration_2017}, we consider the addition of a small linear layer following the final fully-connected layer where either the $4\times 4$ matrix is diagonal (vector calibration) or the $4\times 4$ matrix is allowed to vary freely (matrix calibration). Second, we retrain the final fully-connected layer of the Resnet-34 model on samples from the target domain (fine-tuning). More information on these methods can be found in Appendix~\ref{sec:appendix_adapt}. \\

To investigate the impact of the target domain dataset size, the adaptation techniques were repeated for different SM training sets across a range of sizes. The adaptation process was repeated 10 times for each dataset size with different training data in order to investigate the uncertainty arising from the selection of the training data. For each realization, the training images were randomly selected, without replacement, from the full training set. As a reference, a Resnet-34 model was trained from scratch (ie random initialization) for the largest number of training samples for each SM dataset.

\subsection{Statistical Analysis} \label{sec:evaluation_methods}
To obtain an exam-level assessment, each image within an exam was processed by the DL model and the resulting probabilities were averaged. Several metrics were computed from these average probabilities for the four-class BI-RADS breast density task and the binary dense (BI-RADS C~and~D) versus non-dense (BI-RADS A~and~B) task: (a) accuracy, estimated based on concordance with the original reporting radiologists, (b) the area under the receiver operating characteristic curve (AUC), and (c) Cohen's kappa (https://scikit-learn.org, version~0.20.0). Confidence intervals (CI) were computed by use of non-Studentized pivotal bootstrapping of the test sets for 8000 random samples \cite{carpenter_bootstrap_2000}. For the four-class problem, macroAUC (the average of the four AUC values from the one versus others tasks) and Cohen's kappa with linear weighting are reported. For the binary density tasks, the predicted dense and non-dense probabilities were computed by summing the probabilities for the corresponding BI-RADS density categories. For comparisons of accuracy and Cohen’s kappa before and after adaptation, P-values were calculated using a two-sided z-test with an alpha of 0.05 (https://scipy.org, version~1.1.0; https://www.statsmodels.org, version~0.11.1) \cite{fleiss_large_1969}.

\section{Results}

\subsection{Performance on FFDM Exams} \label{sec:performance_source_domain}

\begin{figure}
    \centering
    \begin{subfigure}[b]{0.5\textwidth}
        \includegraphics[width=\textwidth]{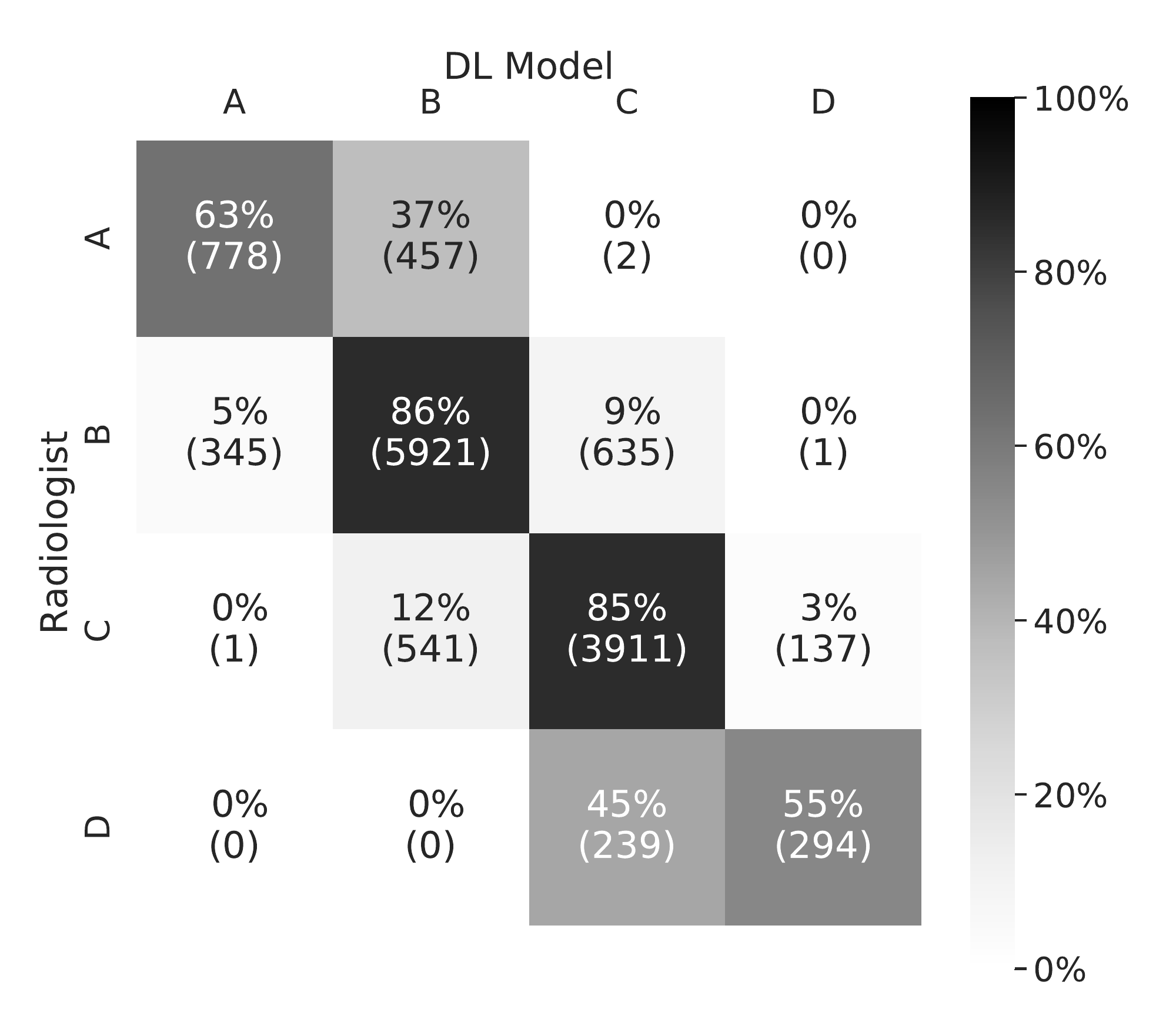}
        \caption{} \label{fig:confusion_matrices_4class_wustl_ffdm}
    \end{subfigure}%
    \begin{subfigure}[b]{0.5\textwidth}
        \includegraphics[width=\textwidth]{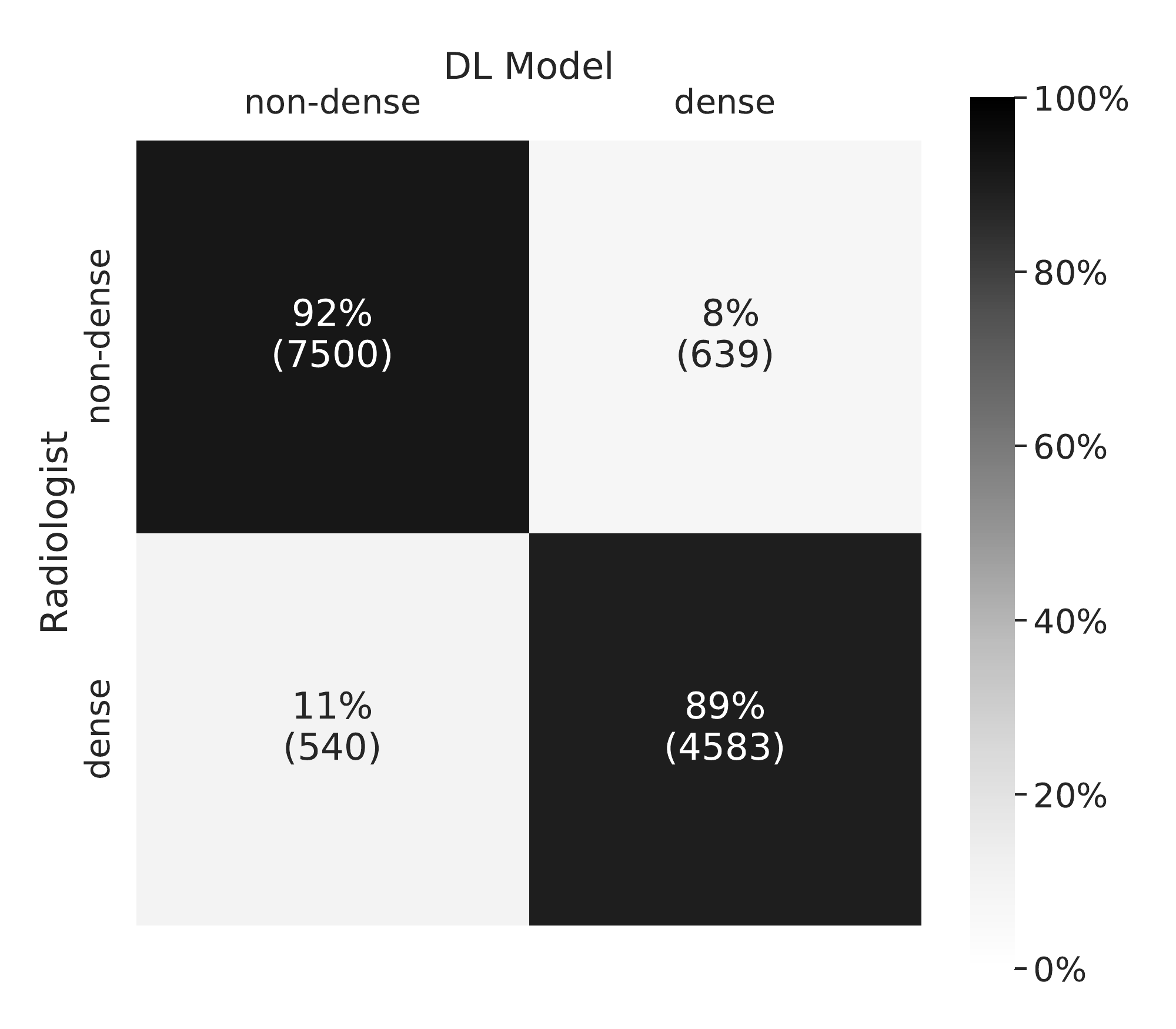}
        \caption{} \label{fig:confusion_matrices_binary_wustl_ffdm}
    \end{subfigure}%
    \caption{Confusion matrices for the A, Breast Imaging Reporting and Data System (BI-RADS) breast density task and the B, binary density task (dense [BI-RADS C and D] vs non-dense [BI-RADS A and B]) evaluated on the full-field digital mammography test set. The number of test samples (exams) within each bin is shown in parentheses. DL = deep learning}
    \label{fig:confusion_matrics_wustl_ffdm}
\end{figure}

The trained model was first evaluated on a large held-out test set of FFDM exams from Site~1 (4960 patients, 13262 exams, 53048 images [mean age, 57 years; age range, 23-97 years]). In this case, the images were from the same institution and of the same image type (FFDM) as employed to train the model. The BI-RADS breast density distribution predicted by the DL model (A, 8.5\%; B, 52.2\%; C, 36.1\%; D, 3.2\%) was similar to that of the original reporting radiologists (A, 9.3\%; B, 52.0\%; C, 34.6\%; D, 4.0\%). A more detailed comparison of the density distributions can be found in Appendix~\ref{sec:appendix_density_dist}. The DL model exhibited close agreement with the radiologists for the BI-RADS breast density task across a variety of performance measures (see Table~\ref{tab:wustl_ffdm_performance}), including accuracy (82.2\% [95\%~CI:~81.6\%,~82.9\%]) and linearly-weighted Cohen's kappa ($\kappa_w$~=~0.75 [95\%~CI:~0.74,~0.76]). A high-level of agreement was also observed for the binary breast density task (accuracy,~91.1\% [95\%~CI:~90.6\%,~91.6\%]; AUC,~0.971, [95\%~CI:~0.968,~0.973], $\kappa$~=~0.81 [95\%~CI:~0.80,~0.82]). As demonstrated by the confusion matrices shown in Figure~\ref{fig:confusion_matrics_wustl_ffdm}, the DL model is rarely off by more than one breast density category (eg calls an extremely dense breast scattered; 0.03\% [4 of 13262]).\\

To place the results in the context of prior work, the performance on the FFDM test set was compared with results evaluated on other large FFDM datasets acquired from academic centers \cite{lehman_mammographic_2019, wu_breast_2017}, with commercial breast density software \cite{brandt_comparison_2016}, and with an estimate of human performance \cite{sprague_variation_2016} (see Table~\ref{tab:wustl_ffdm_performance}). While there are limitations to comparing results evaluated on different datasets with different readers (see the Discussion section), our FFDM DL model appears to offer competitive performance.

\begin{table}[htbp]
    \centering
    \caption{Performance of the Baseline Model on the Test Set for FFDM Exams for both the Four-class BI-RADS Breast Density Task and the Binary Density Task}
    \begin{tabular}{lcccccc}
        \vspace*{1mm}\\
        \toprule
         ~ & \shortstack{4-class\\Accuracy} & \shortstack{4-class\\macroAUC} & \shortstack{4-class\\Linear $\kappa$} & \shortstack{Binary\\Accuracy} & \shortstack{Binary\\AUC} & \shortstack{Binary\\$\kappa$} \\
         \midrule
         Ours & \shortstack{82.2 \\{[81.6, 82.9]}} & \shortstack{0.952 \\{[0.949, 0.954]}} & \shortstack{0.75 \\{[0.74, 0.76]}} & \shortstack{91.1 \\{[90.6, 91.6]}} & \shortstack{0.971 \\{[0.968, 0.973]}} & \shortstack{0.81 \\{[0.80, 0.82]}} \\
         \midrule
         Lehman et al.~\cite{lehman_mammographic_2019} & \shortstack{77 \\{[76, 78]}} & & \shortstack{0.67 \\{[0.66, 0.68]}} & \shortstack{87 \\{[86, 88]}} \\
         \midrule
         Wu et al.~\cite{wu_breast_2017} & 76.7 & 0.916 & & 86.5 & & 0.65 \\
         \midrule
         Volpara v1.5.0~\cite{brandt_comparison_2016} & 57 & & \shortstack{0.57 \\{[0.55, 0.59]}} & 78 & & \shortstack{0.64 \\{[0.61, 0.66]}} \\
         Quantra v2.0~\cite{brandt_comparison_2016} & 56 & & \shortstack{0.46 \\{[0.44, 0.47]}} & 83 & & \shortstack{0.59 \\{[0.57, 0.62]}}\\
         \midrule
         \shortstack{Inter-radiologist\\Agreement \cite{sprague_variation_2016}} & 67.4 & & & 82.8 & & \\
         \bottomrule
    \end{tabular}
    \vspace*{3mm}
    \caption*{Note.--The 95\% confidence intervals are shown in brackets. Binary density task denotes performance of dense (BI-RADS~C~and~D) versus non-dense (BI-RADS~A~and~B). Results from prior studies on automated BI-RADS breast density models are shown evaluated on their respective test sets as points of comparison. An estimate of human performance is provided as a reference. AUC~=~area under the receiver operating characteristic curve, BI-RADS~=~Breast Imaging and Reporting Data System, FFDM~=~full-field digital mammography}
    \label{tab:wustl_ffdm_performance}
\end{table}

\subsection{Exam-level Performance on SM Images}

\subsubsection{Site 1 Results}

\begin{figure}
    \centering
    \begin{subfigure}[b]{0.5\textwidth}
        \includegraphics[width=\textwidth]{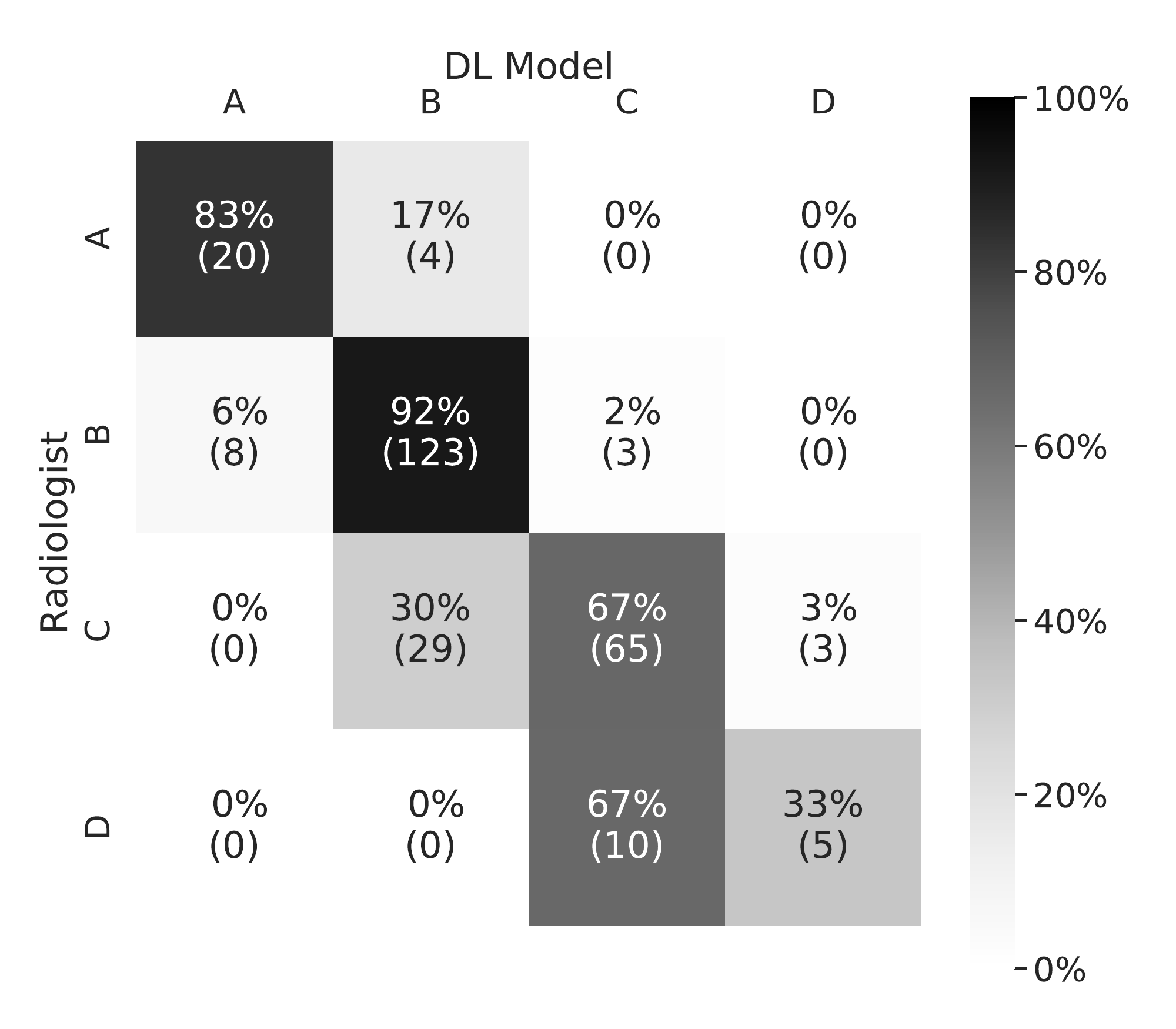}
        \caption{} \label{fig:confusion_matrices_4class_wustl_cview}
    \end{subfigure}%
    \begin{subfigure}[b]{0.5\textwidth}
        \includegraphics[width=\textwidth]{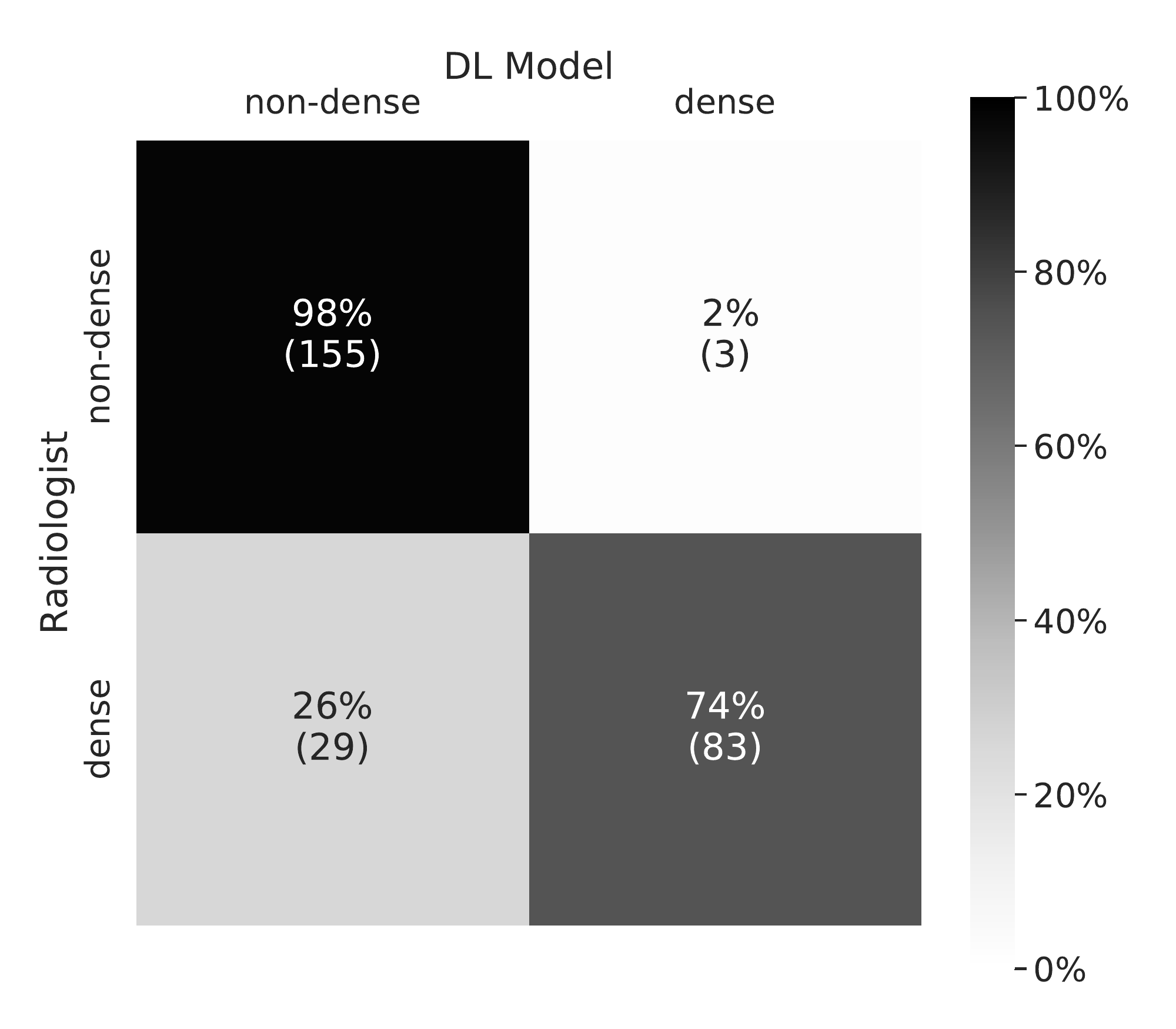}
        \caption{} \label{fig:confusion_matrices_binary_wustl_cview}
    \end{subfigure}\\
    \begin{subfigure}[b]{0.5\textwidth}
        \includegraphics[width=\textwidth]{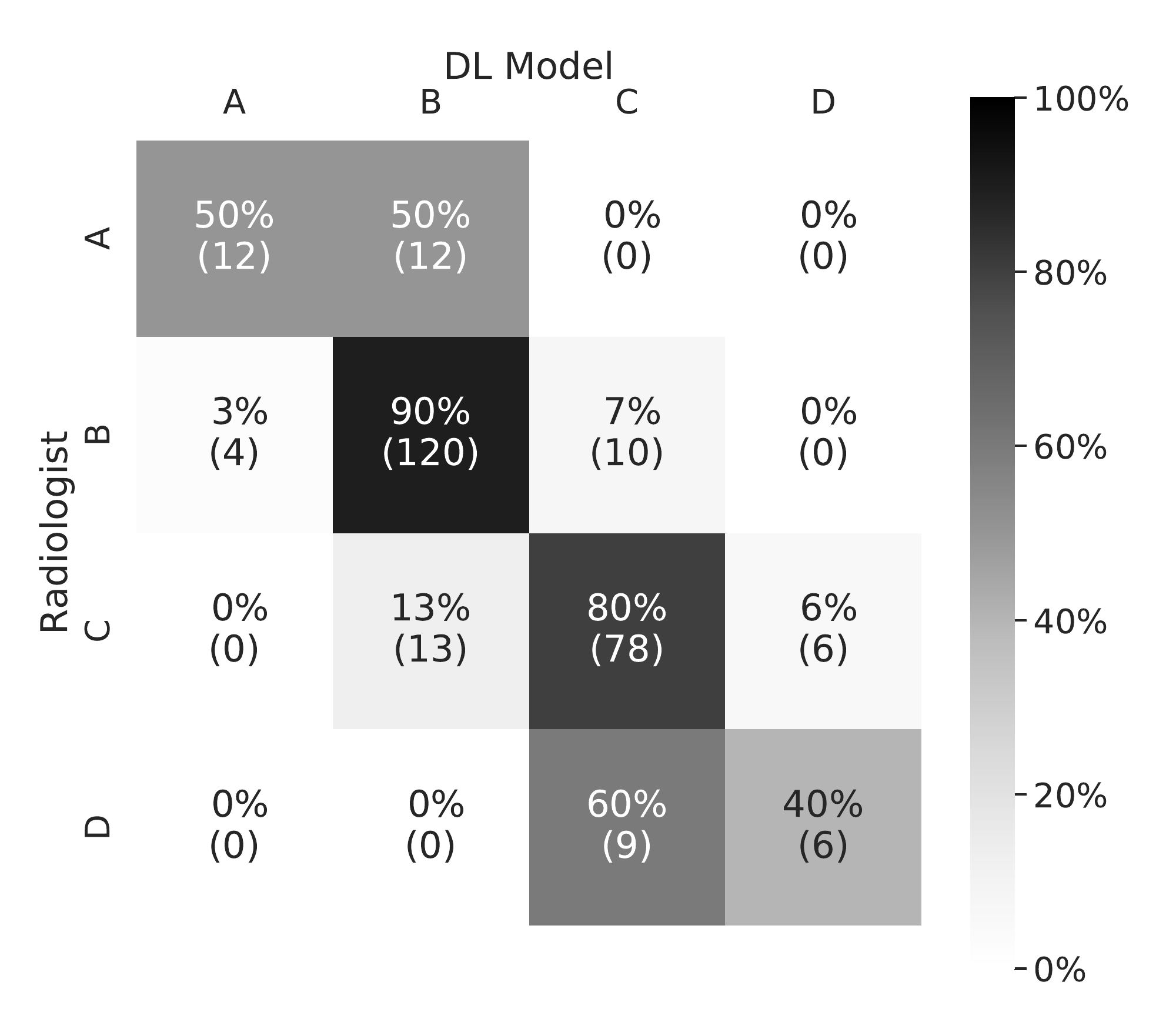}
        \caption{} \label{fig:confusion_matrices_4class_wustl_cview_matcal}
    \end{subfigure}%
    \begin{subfigure}[b]{0.5\textwidth}
        \includegraphics[width=\textwidth]{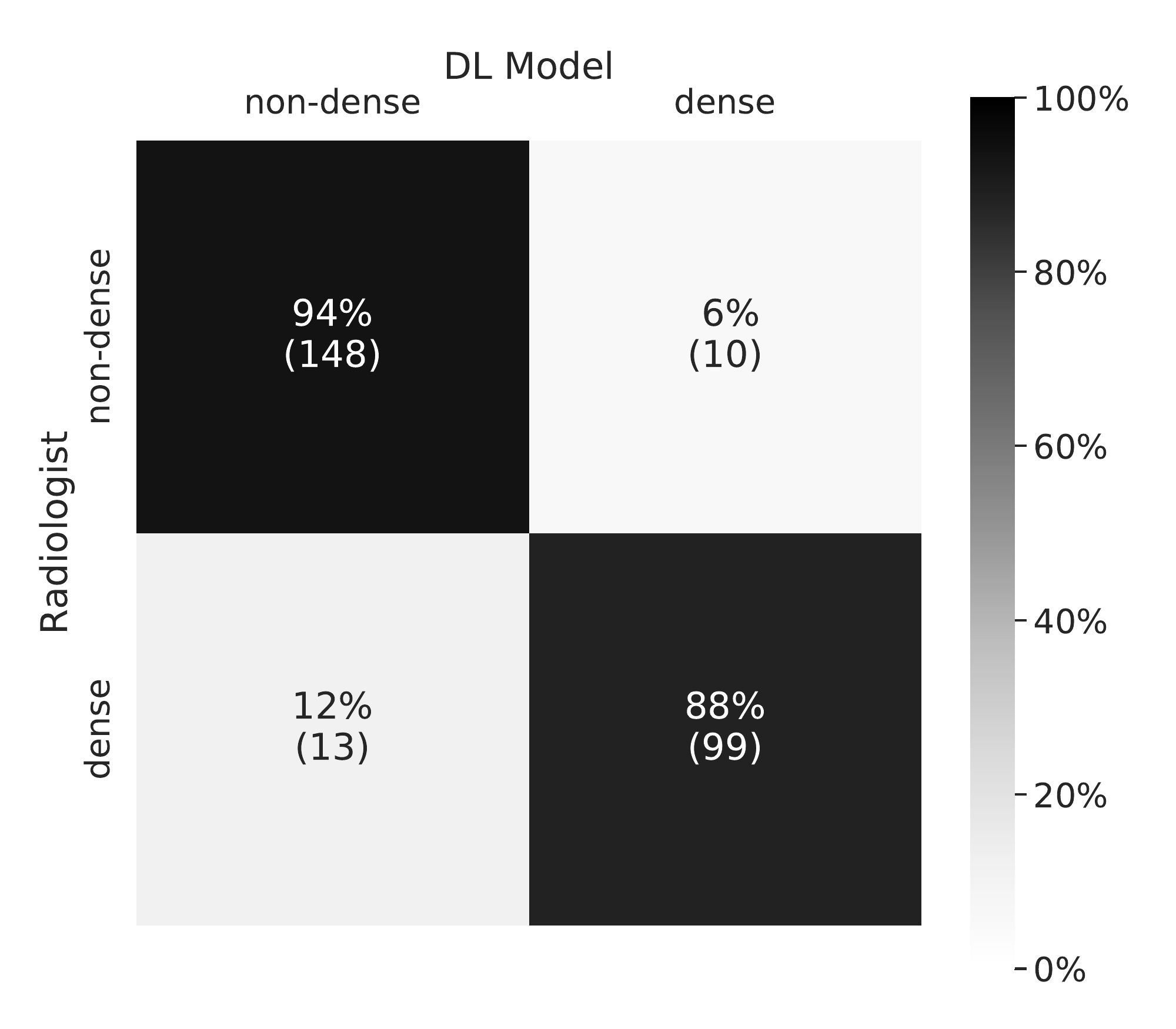}
        \caption{} \label{fig:confusion_matrices_binary_wustl_cview_matcal}
    \end{subfigure}%
    \caption{Confusion matrices, evaluated on the Site 1 synthetic two-dimensional mammography (SM) test set without adaptation, for the A, Breast Imaging Reporting and Data System (BI-RADS) breast density task and B, the binary density task (dense [BI-RADS~C~and~D] vs non-dense [BI-RADS~A~and~B]). Confusion matrices, evaluated on the Site~1 SM test set with adaptation by matrix calibration for 500 Site~1 SM training samples, for the C, BI-RADS breast density task and D, the binary density task (dense vs non-dense). The number of test samples (exams) within each bin is shown in parentheses. DL = deep learning}
    \label{fig:confusion_matrices_wustl_cview}
\end{figure}

Results are first reported for the Site 1 SM test set (270 patients, 270 exams, 1080 images [mean age, 55 years; age range, 28-72 years]) as this avoids any differences that may occur between the two sites. Without adaptation, the model still demonstrates close agreement with the original reporting radiologists for the BI-RADS breast
density task (accuracy, 79\% [95\% CI: 74\%, 84\%]; $\kappa_w$ = 0.71 [95\% CI: 0.64, 0.78]; see Table~\ref{tab:adaptation_performance}). The DL model slightly underestimates breast density for SM images (see Figure~\ref{fig:confusion_matrices_wustl_cview}), producing a BI-RADS breast density distribution (A,~10.4\%; B,~57.8\%; C,~28.9\%; D, 3.0\%) with more non-dense cases and fewer dense cases relative to the radiologists (A,~8.9\%; B,~49.6\%; C,~35.9\%; D,~5.6\%). A more detailed comparison of the density distributions can be found in Appendix~\ref{sec:appendix_density_dist}. Agreement for the binary density task is also quite high without adaptation (accuracy, 88\% [95\% CI: 84\%, 92\%]; $\kappa$~=~0.75 [95\% CI: 0.67, 0.83]; AUC~=~0.97 [95\% CI: 0.96, 0.99]). \\

After adaptation by matrix calibration with 500 Site~1 SM images, the density distribution is slightly more similar to that of the radiologists (A, 5.9\%; B, 53.7\%; C, 35.9\%; D, 4.4\%), while overall agreement is about the same
(accuracy, 80\% [95\% CI: 76\%, 85\%], P = .75; $\kappa_w$~=~0.72 [95\% CI: 0.66, 0.79], P = .80). Accuracy for the two dense classes is improved at the expense of the two non-dense classes (see Figure~\ref{fig:confusion_matrices_wustl_cview}). A larger, though not statistically
significant, improvement is seen for the binary density task, where Cohen's kappa rose from 0.75 [95\% CI: 0.67, 0.83) to 0.82 [95\% CI: 0.76, 0.90], P = .16 (accuracy, 91\% [95\% CI: 88\%, 95\%], P = .20).

\subsubsection{Site 2 Results}
\begin{figure}
    \centering
    \begin{subfigure}[b]{0.5\textwidth}
        \includegraphics[width=\textwidth]{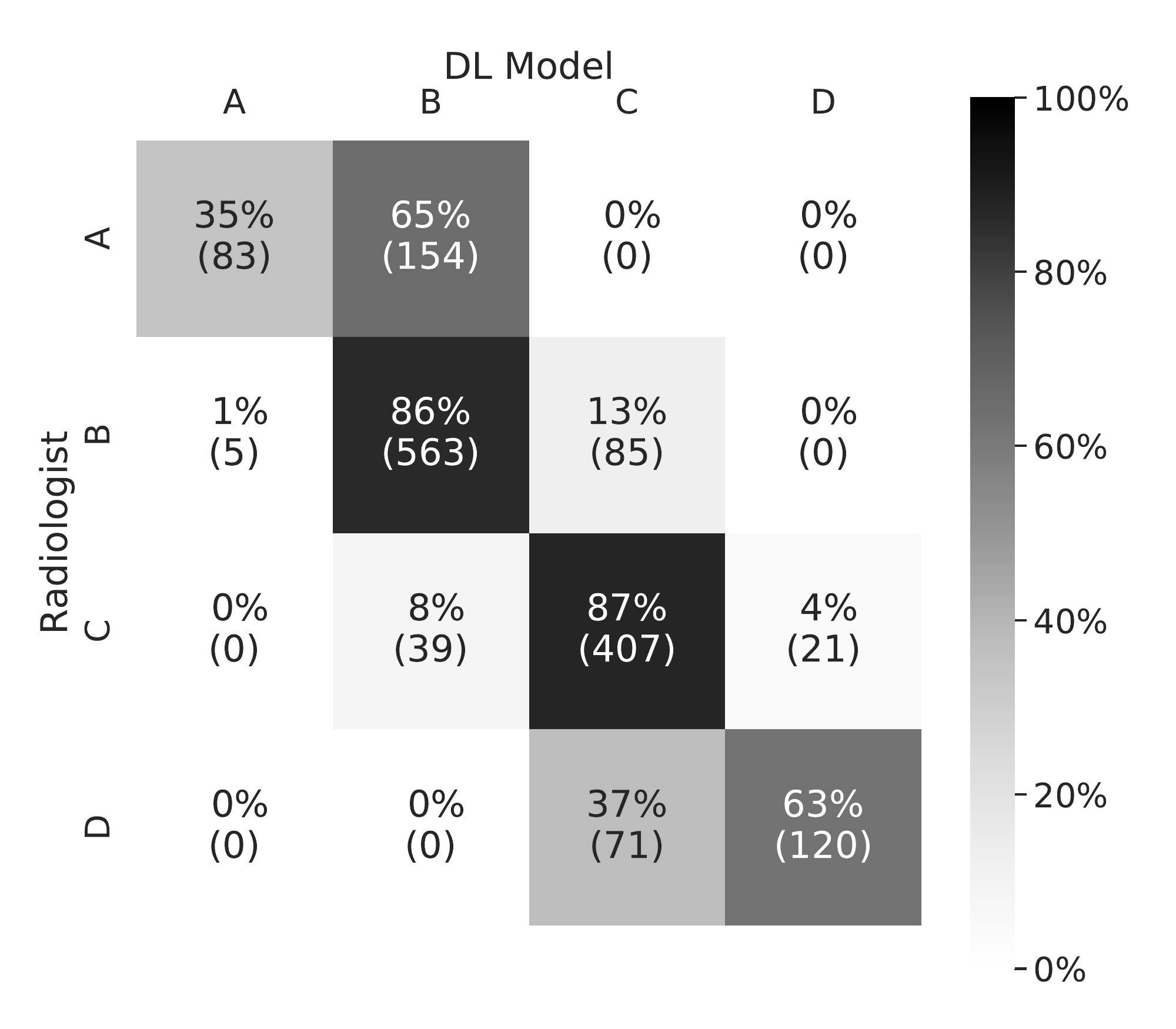}
        \caption{} \label{fig:confusion_matrices_4class_pdi_cview}
    \end{subfigure}%
    \begin{subfigure}[b]{0.5\textwidth}
        \includegraphics[width=\textwidth]{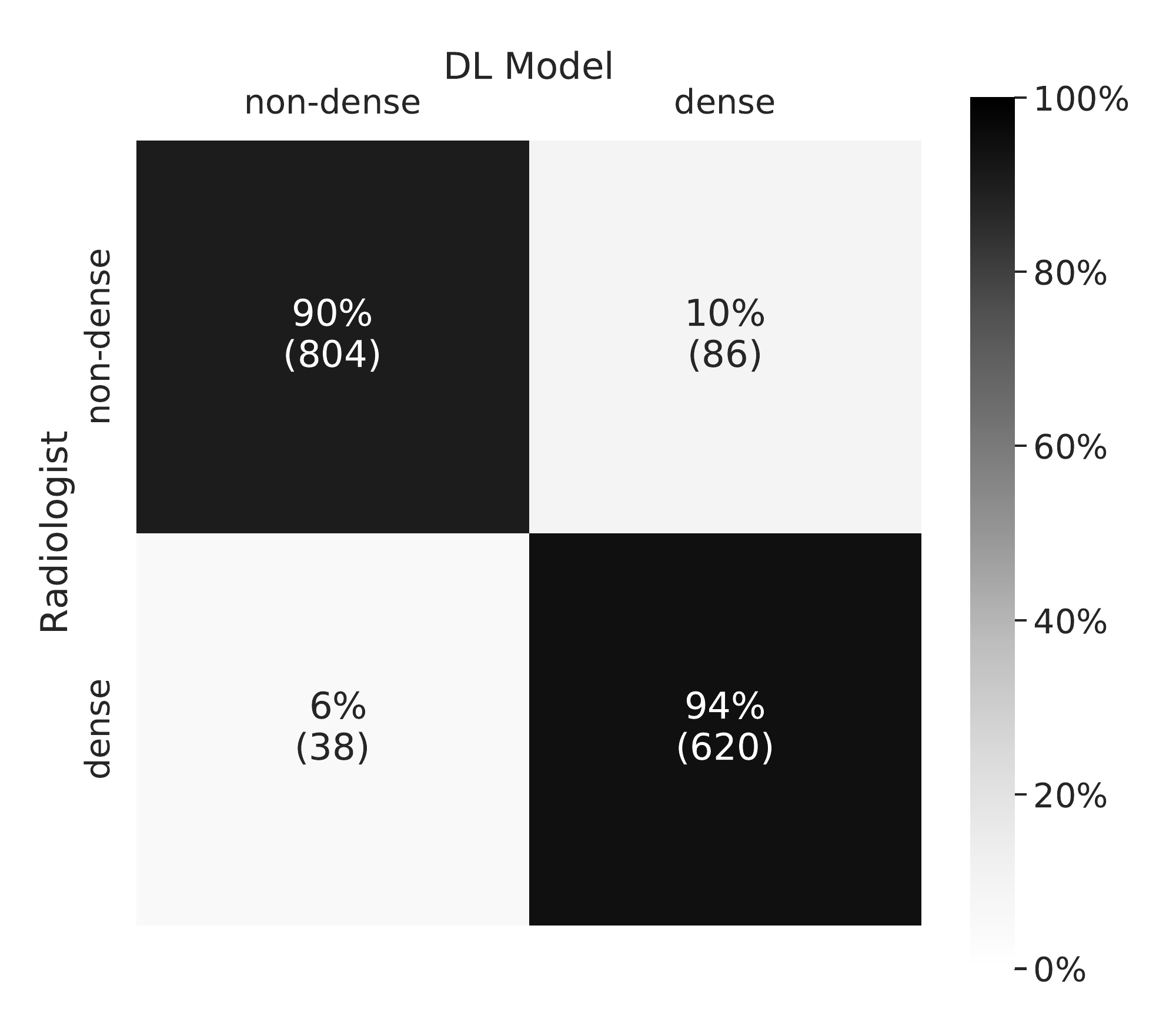}
        \caption{} \label{fig:confusion_matrices_binary_pdi_cview}
    \end{subfigure}\\
    \begin{subfigure}[b]{0.5\textwidth}
        \includegraphics[width=\textwidth]{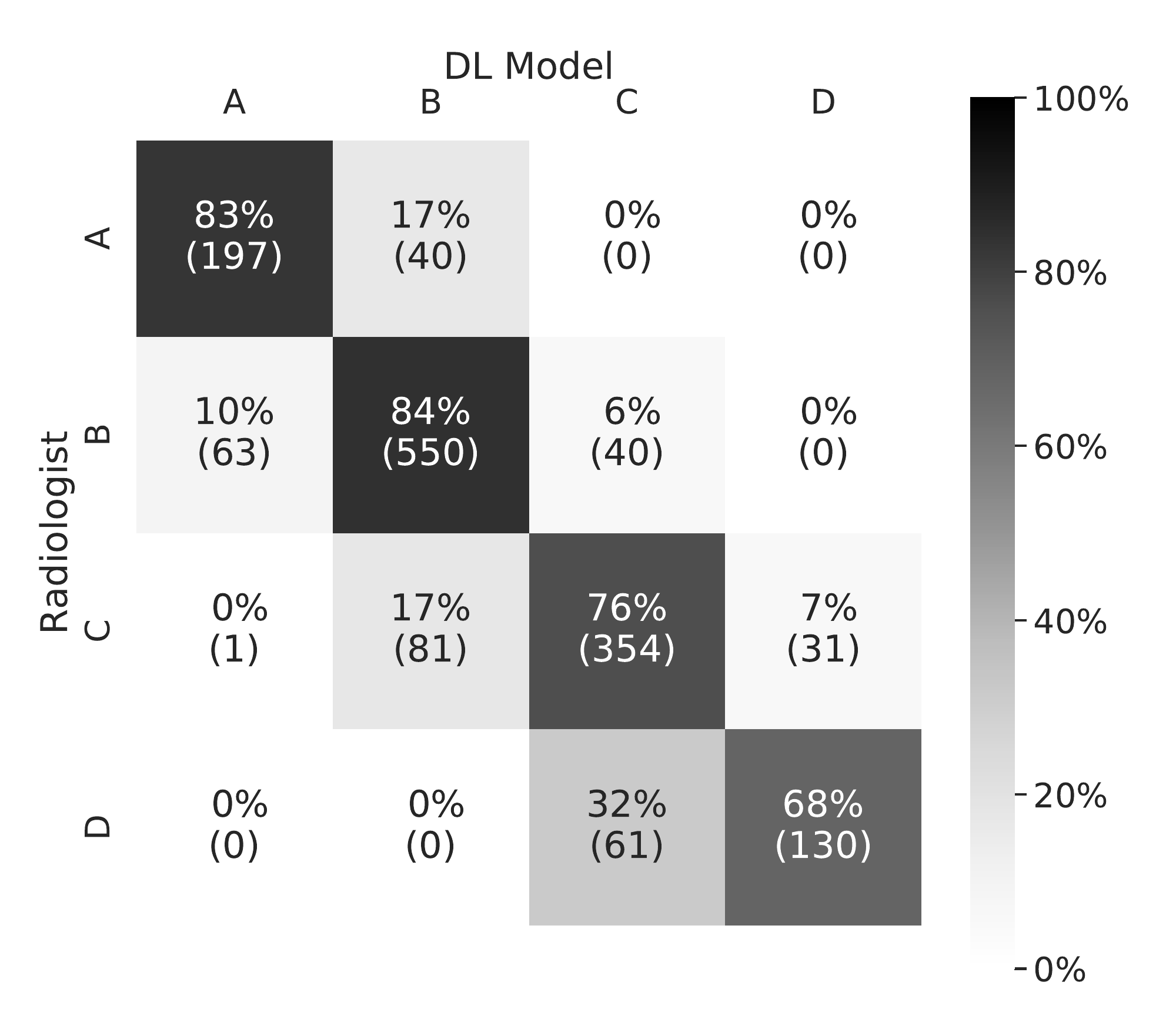}
        \caption{} \label{fig:confusion_matrices_4class_pdi_cview_matcal}
    \end{subfigure}%
    \begin{subfigure}[b]{0.5\textwidth}
        \includegraphics[width=\textwidth]{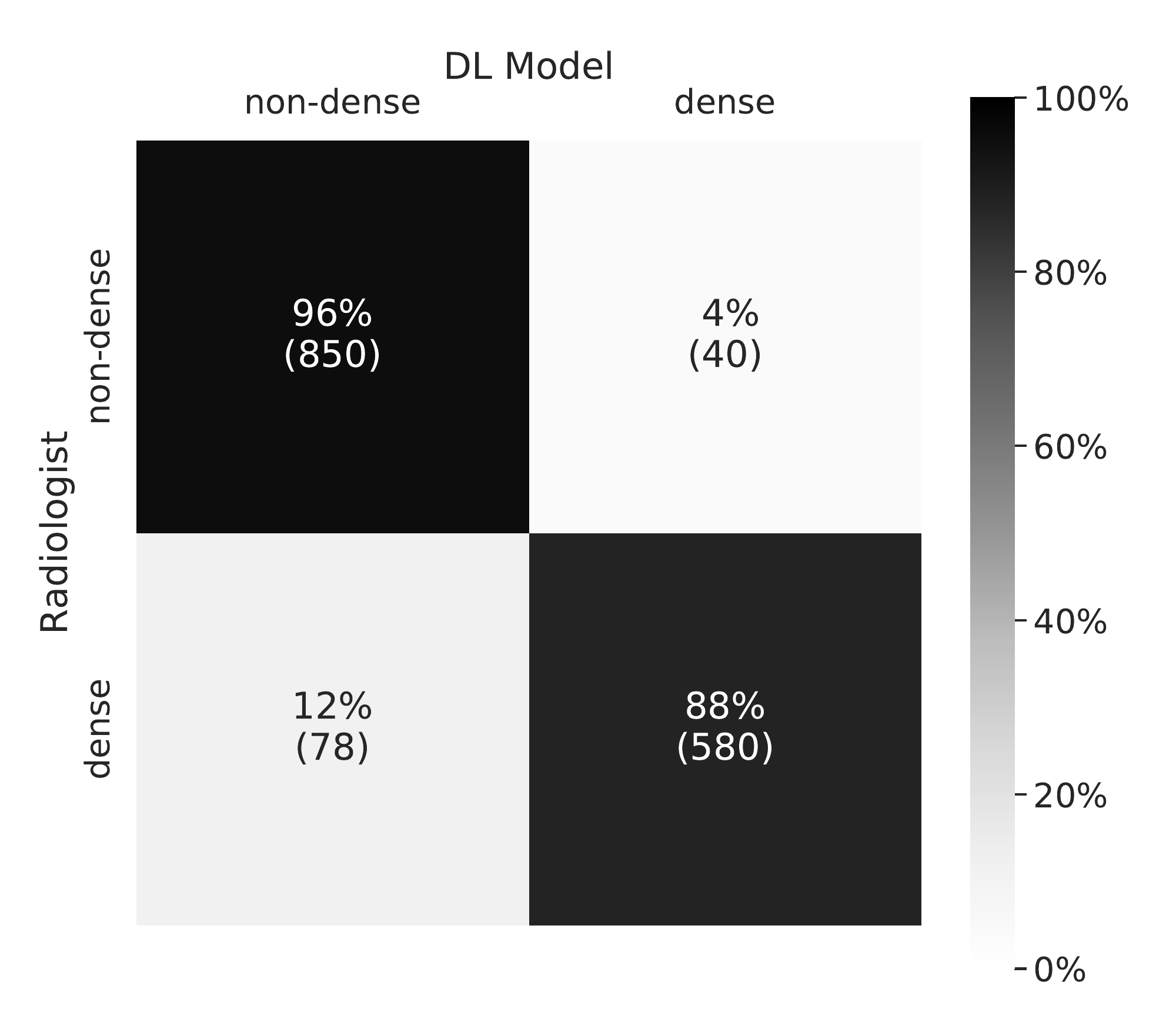}
        \caption{} \label{fig:confusion_matrices_binary_pdi_cview_matcal}
    \end{subfigure}%
    \caption{Confusion matrices, evaluated on the Site~2 synthetic two-dimensional mammography (SM) test set without adaptation, for the A, Breast Imaging Reporting and Data System (BI-RADS) breast density task and B, the binary density task (dense [BI-RADS~C~and~D] vs non-dense [BI-RADS~A~and~B]). Confusion matrices, evaluated on the Site~2 synthetic two-dimensional mammography (SM) test set with adaptation by matrix calibration for 500 Site~2 SM training samples, for the C, BI-RADS breast density task and D, the binary density task (dense vs. nondense). Number of test samples (exams) within each bin are shown in parentheses.}
    \label{fig:confusion_matrices_pdi_cview}
\end{figure}

Close agreement between the DL model and the original reporting radiologists was also observed for the Site~2 SM test set (744 patients, 1548 exams, 6192 images [mean age, 55 years; age range, 30-92 years]) without adaptation (accuracy, 76\% [95\%~CI: 74\%, 78\%]; $\kappa_w$~=~0.72 [95\%~CI: 0.70, 0.75]; see Table~\ref{tab:adaptation_performance}). The BI-RADS breast density distribution predicted by the DL model (A,~5.7\%; B,~48.8\%; C,~36.4\%; D,~9.1\%) was similar to the distributions found in the Site~1 datasets. The predicted density distribution does not appear to be skewed towards low density estimates as seen for Site~1 (see Figure~\ref{fig:confusion_matrices_pdi_cview}). Agreement for the binary density task was especially strong (accuracy, 92\% [95\%~CI: 91\%, 93\%]; $\kappa$~=~0.84 [95\%~CI: 0.81, 0.87]; AUC, 0.980 [95\%~CI: 0.976, 0.986]). \\

With adaptation by matrix calibration with 500 Site~2 SM training samples, performance for the BI-RADS breast density task on the Site~2 SM dataset substantially improved (accuracy = 80\% [95\% CI: 78\%, 82\%], P $<$ .001; $\kappa_w$~=~0.79 [95\% CI: 0.76, 0.81], P $<$ .001). After adaptation, the predicted BI-RADS breast density distribution (A,~16.9\%; B,~43.3\%; C,~29.4\%; D,~10.4\%) was more similar to that of the radiologists (A,~15.3\%; B,~42.2\%; C,~30.2\%; D,~12.3\%). Less improvement was seen for the binary breast density task (accuracy, 92\% [95\% CI: 91\%, 94\%], P~=~.69; $\kappa$~=~0.84 [95\% CI: 0.82, 0.87], P = .79).

\begin{table}[htbp]
    \centering
    \caption{Performance of the Proposed Approaches for Adapting a DL Model Trained on one Dataset to another with only 500 SM Images}
    \begin{tabular}{llcccccc}  
        \vspace*{1mm}\\
        \toprule
         Datasets & Methods & \shortstack{4-class\\Accuracy} & \shortstack{4-class\\macroAUC} & \shortstack{4-class\\Linear $\kappa$} & \shortstack{Binary\\Accuracy} & \shortstack{Binary\\AUC} & \shortstack{Binary\\$\kappa$} \\ 
         \midrule
         FFDM & & 82.2 & 0.952 & 0.75 & 91.1 & 0.971 & 0.81\\
         \midrule
         \multicolumn{2}{l}{FFDM $\rightarrow$ Site~1} & & & & & &  \\
         \midrule
           & None & \shortstack{79 \\{[74, 84]}} & \shortstack{0.94 \\{[0.93, 0.96]}} & \shortstack{0.71 \\{[0.64, 0.78]}} & \shortstack{88 \\{[84, 92]}} & \shortstack{0.97 \\{[0.96, 0.99]}} & \shortstack{0.75 \\{[0.67, 0.83]}} \\
          & Vector & \shortstack{81 \\{[77, 86]}} & \shortstack{0.95 \\{[0.94, 0.97]}} & \shortstack{0.73 \\{[0.67, 0.80]}} & \shortstack{90 \\{[87, 94]}} & \shortstack{0.97 \\{[0.96, 0.99]}} & \shortstack{0.80 \\{[0.73, 0.88]}}\\
          & Matrix & \shortstack{80 \\{[76, 85]}} & \shortstack{0.95 \\{[0.94, 0.97]}} & \shortstack{0.72 \\{[0.66, 0.79]}} & \shortstack{91 \\{[88, 95]}} & \shortstack{0.97 \\{[0.96, 0.99]}} & \shortstack{0.82 \\{[0.76, 0.90]}}\\
          & Fine-tune & \shortstack{81 \\{[76, 86]}} & \shortstack{0.95 \\{[0.94, 0.97]}} & \shortstack{0.73 \\{[0.67, 0.80]}} & \shortstack{90 \\{[87, 94]}} & \shortstack{0.97 \\{[0.95, 0.99]}} & \shortstack{0.80 \\{[0.73, 0.88]}}\\
         \midrule
         \multicolumn{2}{l}{FFDM $\rightarrow$ Site~2} & & & & & & \\
         \midrule
         & None & \shortstack{76 \\{[74, 78]}} & \shortstack{0.944 \\{[0.938, 0.951]}} & \shortstack{0.72 \\{[0.70, 0.75]}} & \shortstack{92 \\{[91, 93]}} & \shortstack{0.980 \\{[0.976, 0.986]}} & \shortstack{0.84 \\{[0.81, 0.87]}} \\
          & Vector & \shortstack{79 \\{[77, 81]}} & \shortstack{0.954 \\{[0.949, 0.961]}} & \shortstack{0.78 \\{[0.76, 0.80]}} & \shortstack{92 \\{[91, 93]}} & \shortstack{0.979 \\{[0.974, 0.985]}} & \shortstack{0.83 \\{[0.80, 0.86]}}\\
          & Matrix & \shortstack{80 \\{[78, 82]}} & \shortstack{0.956 \\{[0.950, 0.963]}} & \shortstack{0.79 \\{[0.76, 0.81]}} & \shortstack{92 \\{[91, 94]}} & \shortstack{0.983 \\{[0.978, 0.988]}} & \shortstack{0.84 \\{[0.82, 0.87]}}\\
          & Fine-tune & \shortstack{80 \\{[78, 82]}} & \shortstack{0.957 \\{[0.952, 0.964]}} & \shortstack{0.79 \\{[0.77, 0.81]}} & \shortstack{93 \\{[92, 94]}} & \shortstack{0.984 \\{[0.979, 0.988]}} & \shortstack{0.85 \\{[0.83, 0.88]}}\\
         \bottomrule
    \end{tabular}
    \caption*{Note.--The performance of the model trained from scratch on the full-field digital mammography dataset (672 thousand training samples) and evaluated on its test set is also shown as a reference. 95\% confidence intervals, computed by bootstrapping over the test sets, are given in brackets. AUC~=~area under the receiver operating characteristic curve, DL~=~deep learning, FFDM~=~full-field digital mammography}
    \label{tab:adaptation_performance}
\end{table}

\subsubsection{Impact of Dataset Size on Adaptation}

\begin{figure}
    \centering
    \begin{subfigure}[b]{0.5\textwidth}
        \includegraphics[width=\textwidth]{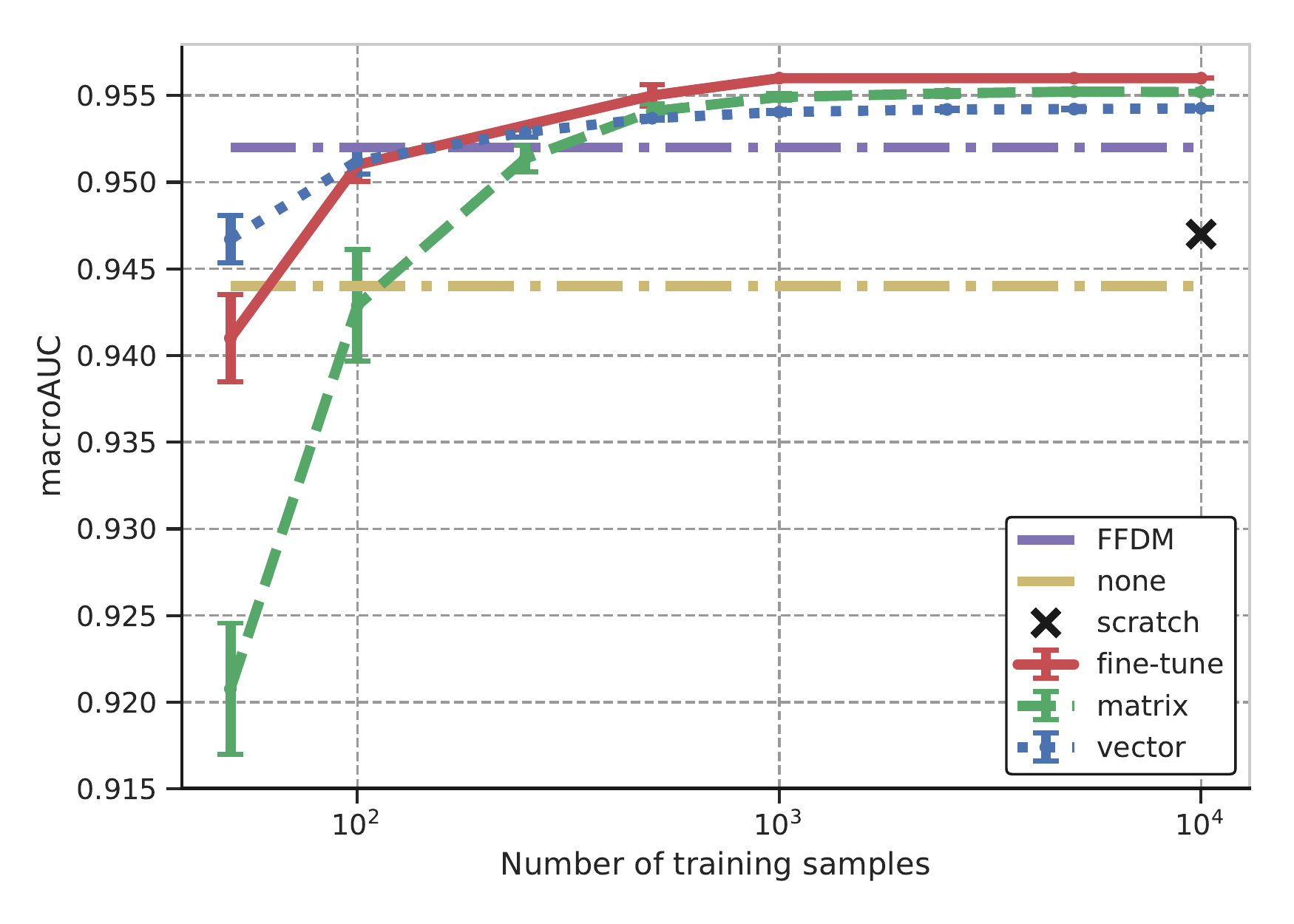}
        \caption{}
    \end{subfigure}%
    \begin{subfigure}[b]{0.5\textwidth}
        \includegraphics[width=\textwidth]{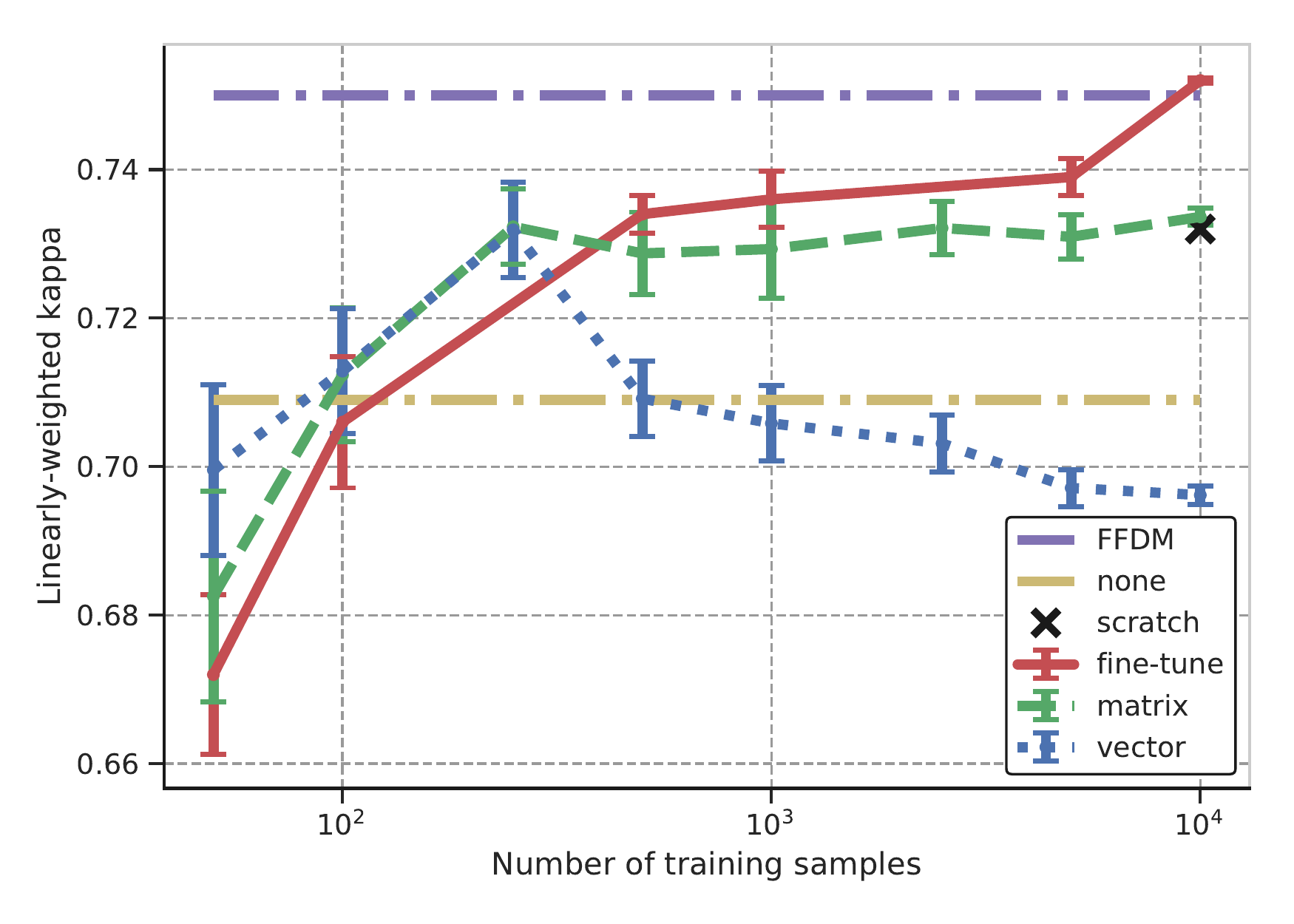}
        \caption{}
    \end{subfigure}\\
    \begin{subfigure}[b]{0.5\textwidth}
        \includegraphics[width=\textwidth]{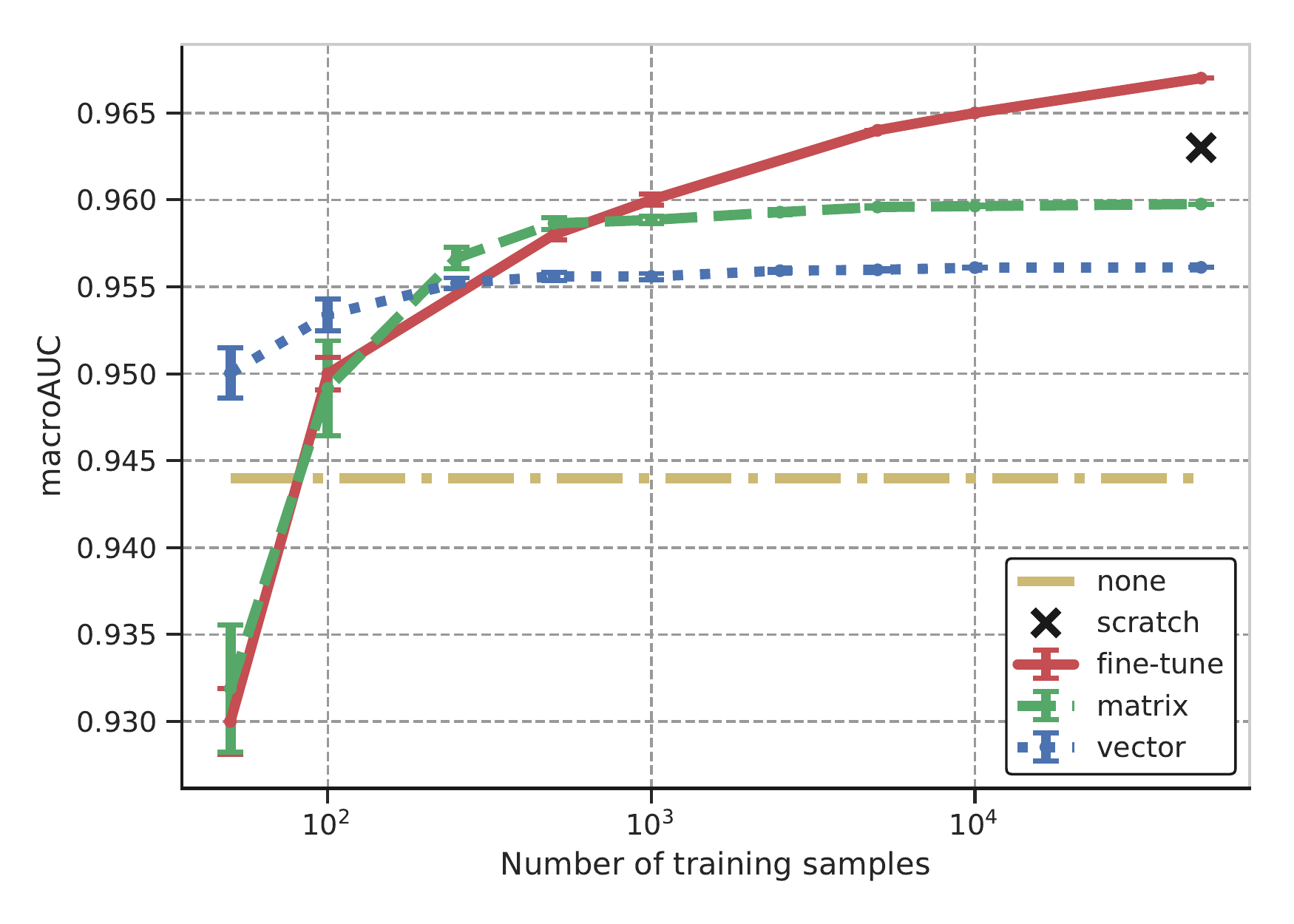}
        \caption{}
    \end{subfigure}%
    \begin{subfigure}[b]{0.5\textwidth}
        \includegraphics[width=\textwidth]{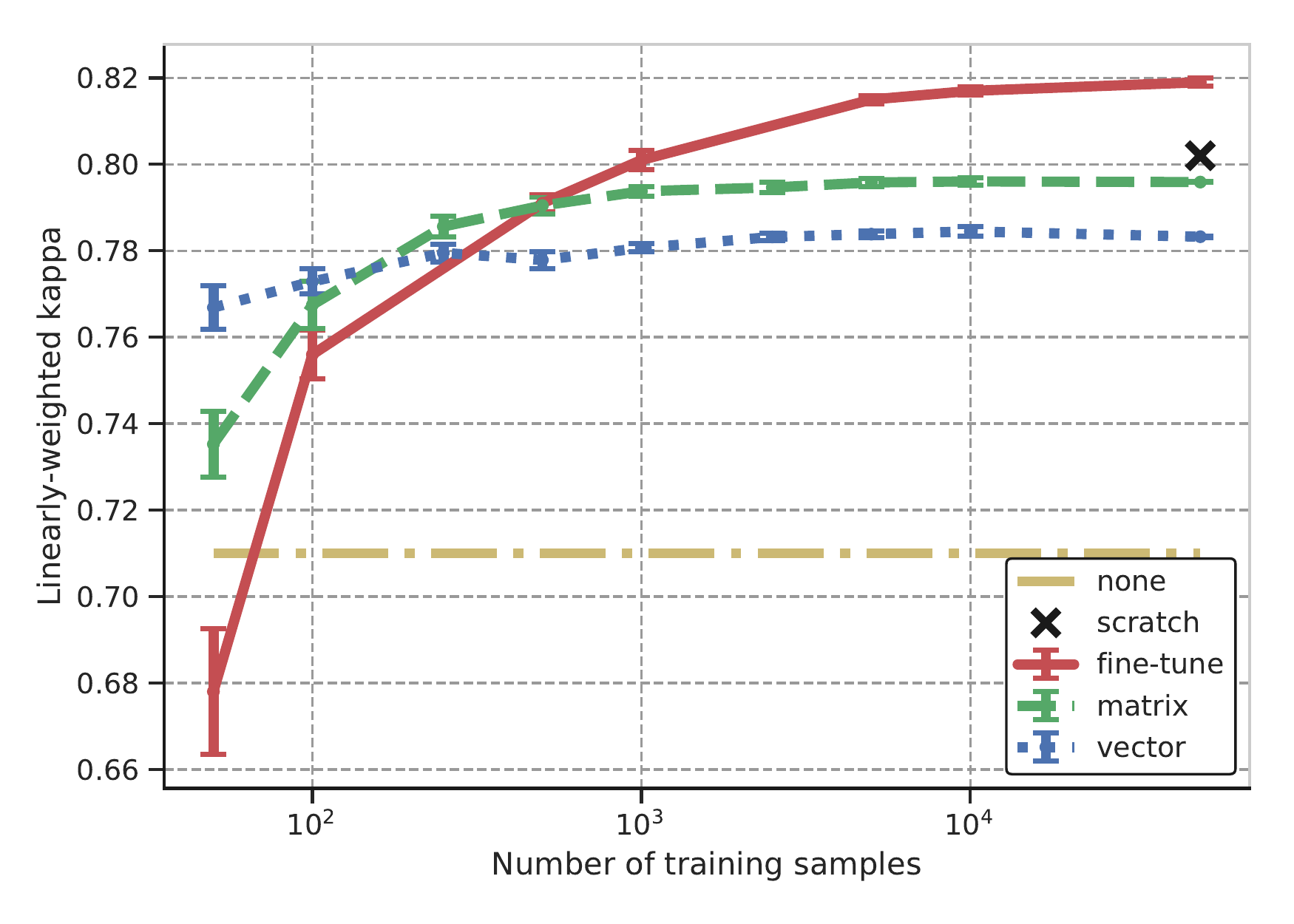}
        \caption{}
    \end{subfigure}%
    \caption{Impact of the number of site-specific training samples in the target domain on the performance of the adapted model for the Site~1 synthetic 2D mammography (SM) test set, as measured by A, macroAUC and B, linearly weighted Cohen's kappa, and for the Site~2 SM test set, as measured by C, macroAUC and D, linearly weighted Cohen's kappa. Results are shown for vector and matrix calibration, and retraining the last fully connected layer (fine-tuning). Error bars indicate the standard error of the mean computed over 10 random realizations of the training data. Performance prior to adaptation (none) and training from scratch are shown as references. For the Site~1 SM studies, the full-field digital mammography (FFDM) performance serves as an additional reference. Note that each graph is shown with its own full dynamic range in order to facilitate comparison of the different adaptation methods for a given metric and dataset.}
    \label{fig:adaptation_data_size}
\end{figure}

The preferred adaptation method will depend on the number of training samples available for the adaptation, with more training samples benefiting methods with more parameters. Figure~\ref{fig:adaptation_data_size} shows the impact of the amount of training data on the performance of the adaptation methods, as measured by linearly weighted Cohen's kappa and macroAUC, for both the Site~1 and Site~2 SM datasets. Each adaptation method has a range of number of samples where it offers the best performance, with the regions ordered by the corresponding number of
parameters for the adaptation methods (vector calibration, 8 parameters; matrix calibration, 20; fine-tuning, 2052). This demonstrates the trade-off between the performance of the adaptation method and the amount of new training data that must be acquired. When the number of training samples is very small (eg $<$ 100 images), some adaptation methods negatively impact performance. Even at the largest dataset sizes, the amount of training data was too limited for the Resnet-34 model trained from scratch on SM images to exceed the performance of the models adapted from FFDM.

\section{Discussion} \label{sec:discussion}
BI-RADS breast density can be an important indicator of breast cancer risk and radiologist sensitivity, but intra- and inter-reader variability may limit the effectiveness of this measure. DL models for estimating breast density can reduce this variability while still providing accurate assessments. However, in order to serve as a useful clinical tool, DL models need to demonstrate that they can be applied to DBT exams and generalize across institutions. To overcome the limited training data for DBT exams, a DL model was trained on a large set of FFDM images. The model showed close agreement with the radiologists reported BI-RADS breast density for a test set of FFDM images (Site~1: $\kappa_w$~=~0.75 [95\%~CI:~0.74,~0.76]) and for two datasets of SM images, which are generated as part of DBT exams (Site~1: $\kappa_w$~=~0.71 [95\%~CI:~0.64,~0.78]; Site~2: $\kappa_w$~=~0.72 [95\%~CI:~0.70,~0.75]). The strong performance on the SM datasets from different institutions suggests that the DL model may generalize to DBT exams and multiple sites. Further adaptation of the model for the SM datasets led to no improvement for Site~1 ($\kappa_w$~=~0.72 [95\%~CI:~0.66,~0.79]) and a more substantial improvement for Site~2 ($\kappa_w$~=~0.79 [95\%~CI:~0.76,~0.81]). The investigation of the impact of dataset size suggests that these adaptation methods could serve as practical approaches for adapting DL models if a model must be updated to account for site-specific differences. \\

When radiologists' assessments are accepted as the ground truth, inter-reader variability may limit the performance that can be achieved for a given dataset. For example, the performance obtained on the Site~2 SM dataset following adaptation was higher than that obtained on the FFDM dataset used to train the model. This is likely a result of more consistency in the ground truth labels for the Site~2 SM dataset due to over 80\% of the exams having been read by two readers.\\

Unlike previous studies, our BI-RADS breast density DL model was evaluated on SM images from DBT exams and on data from multiple institutions. Further, when evaluated on the FFDM images, the model appeared to offer competitive performance to previous DL models and commercial breast density software ($\kappa_w$~=~0.75 [95\%~CI:~0.74,~0.76] vs Lehman et al, 0.67 [95\%~CI:~0.66,~0.68]; Volpara, 0.57 [95\%~CI:~0.55,~0.59]; Quantra 0.46 [95\% CI: 0.44, 0.47]) \cite{lehman_mammographic_2019, brandt_comparison_2016}. Estimates of the model performance appear comparable, or even superior, to previous estimates of inter-radiologist variability for the BI-RADS breast density task \cite{sprague_variation_2016}. For each automated breast density method, results are reported on their respective test sets, which may be more or less challenging due to varying levels of inter-reader variability or other factors. Additionally, many performance metrics, such as accuracy and Cohen’s kappa, depend on the prevalence of the BI-RADS breast density categories. Whether the model is evaluated against the assessments of individual radiologists or a consensus of multiple radiologists can also impact the apparent performance of the model. The provided performance numbers from our work and prior work are based on the assessments of individual radiologists. \\

Other measures of breast density, such as volumetric breast density, have been previously estimated by automated software for DBT exams \cite{tagliafico_estimation_2013, pertuz_fully_2015, fornvik_comparison_2019}. Thresholds can be chosen to translate these measures to BI-RADS breast density, but this may result in lower levels of agreement than direct estimation of BI-RADS breast density (eg $\kappa_w$~=~0.47~\cite{fornvik_comparison_2019}). Here, BI-RADS breast density is estimated from 2D SM images instead of the 3D tomosynthesis
volumes as this simplifies transfer learning from the FFDM images. This is also the manner in which breast radiologists assess density for DBT exams. \\

This study has several limitations. First, the proposed domain adaptation approaches may be less effective when the differences between domains are larger. In this work, adaptation is from two types of mammography images produced by the same manufacturer. Second, the FFDM data from Site~1 was collected over a time period covering
the transition from BI-RADS version~4 to BI-RADS version~5, during which the criteria for assessing BI-RADS breast density changed. Third, the test set includes multiple exams from the same patient, which could have led to underestimation of the variance for the given performance measures. Fourth, the reference standard was breast
density assessed by the original interpreting radiologist, which is known to have inter- and intra-reader variation. Fifth, when a DL model is adapted to a new institution, adjustments may be made for differences in image content,
patient demographics, or the interpreting radiologists. This last adjustment may result in a degree of inter-reader variability between the original and adapted DL models, though likely lower than the individual inter-reader variability if the model learns the consensus of each group of radiologists. As a result, the improved performance
following adaptation for the Site~2 SM dataset could be due to differences in patient demographics or radiologist assessment practices compared with the FFDM dataset. The weaker improvement for the Site~1 SM dataset could be due to similarities in these same factors. \\

Still, the broad use of BI-RADS breast density DL models holds great promise for improving clinical care. The success of the DL model without adaptation suggests that the features learned by the model are largely applicable to both FFDM images and SM images from digital breast tomosynthesis exams as well as to different readers and institutions. A BI-RADS breast density DL model that can generalize across sites and image types could lead to rapid and more consistent estimates of breast density for women.

\appendix
\renewcommand\thefigure{\thesection.\arabic{figure}}
\renewcommand\thetable{\thesection.\arabic{table}}
\setcounter{figure}{0} 
\setcounter{table}{0} 

\section{Training Procedure} \label{sec:training_procedure}

The deep learning (DL) model, described in the Materials and Methods Section, was a pre-activation Resnet-34 network, where the batch normalization layers were replaced with group normalization layers \cite{he_identity_2016, he_deep_2016,wu_group_2018}. It was trained using the full-field digital mammography (FFDM) dataset (see Table~\ref{tab:wustl_datasets}) by use of the Adam optimizer \cite{kingma_adam:_2015} with a learning rate of $10^{-4}$ and a weight decay of $10^{-3}$. Weight decay not was applied to the parameters belonging to the normalization layers. The input was resized to 416x320 pixels and the pixel intensity values were normalized so that the grayscale window denoted in the Digital Imaging and Communications in Medicine (DICOM) header ranged from 0.0 to 1.0. No additional preprocessing was performed. Training was performed using mixed precision \cite{micikevicius_mixed_2017} and gradient checkpointing \cite{chen_training_2016}  with batch sizes of 256 distributed across two NVIDIA GTX 1080 Ti graphics processing units (Santa Clara, CA). Each batch was sampled such that the probability of selecting a BI-RADS~B or BI-RADS~C sample was four times that of selecting a BI-RADS~A or BI-RADS~D sample, which roughly corresponds to the distribution of densities observed nationally in the United States \cite{lehman_national_2017}. Horizontal and vertical flipping were employed for data augmentation. In order to obtain more frequent information on the training progress, epochs were capped at 100k samples compared with a total training set size of over 672k samples. The model was trained for 100 such epochs. Results are reported for the epoch that had the lowest cross entropy loss on the validation set, which occurred after 93 epochs.\\

Training from scratch on the synthetic 2D mammography (SM) datasets was performed following the same procedure as for the base model. For training from scratch, the size of an epoch was set to the number of training samples.

\section{Adaptation Methods}\label{sec:appendix_adapt}
Three methods were employed to adapt the DL model from FFDM to SM and from Site 1 to Site 2: 1) vector calibration, 2) matrix calibration, and 3) fine-tuning.\\

The two calibration methods were originally proposed by Guo et al. for the task for calibration and have been repurposed here for adaptation \cite{guo_calibration_2017}. In both methods, a linear operator is applied to the logits produced by the existing model. The parameters associated with the linear operator are learned as part of the adaptation process as described below. The updated probabilities predicted by the model are, then, given by:
\begin{align}
    \mathbf{p}~=~\sigma\left(\mathbf{A} \mathbf{z} + \mathbf{b}\right),
\end{align}
where $\mathbf{z}$ is the logits, $\mathbf{A}$ and $\mathbf{b}$ are the weights and bias associated with the linear operator, $\sigma$ is the softmax operator, and $\mathbf{p}$ is the updated probabilities that an image belongs to each of the BI-RADS breast density categories. For vector calibration, $\mathbf{A}$ must be a diagonal matrix, while for matrix calibration, no restrictions are placed on $\mathbf{A}$. The parameters for the vector and matrix calibration methods were chosen by minimizing a cross-entropy loss function by use of the BFGS optimization method (https://scipy.org, version 1.1.0). The parameters were initialized such that the linear layer corresponded to the identity transformation. Training was stopped when the $\ell_2$-norm of the gradient was less than $10^{-6}$ or when the number of iterations exceeded 500. \\

For the fine-tuning method, no new layers or weights are introduced. Instead, a small portion of the network is re-trained. In our case, only the last fully-connected layer is updated. For all layers, the model weights were initialized with the weights resulting from training on the FFDM dataset. Retraining the last fully-connected layer
for the fine-tuning method was performed by use of the Adam optimizer with a learning rate of $10^{-4}$ and weight decay of $10^{-5}$. The batch size was set to 64. The fully-connected layer was trained from random initialization for 100 epochs and results were reported for the epoch with the lowest validation cross entropy loss. For fine-tuning, the size of an epoch was set to the number of training samples.

\section{Density Distributions}\label{sec:appendix_density_dist}
The similarity between the DL model and radiologist density distributions is evaluated by use of several statistical techniques. Statistical significance for the difference between the radiologist and DL model density distributions is computed using a Pearson’s chi squared test. The similarity between the two distributions is also estimated using the Kullback-Liebler (KL) divergence where the radiologist distribution serves as the reference distribution. The 95\% confidence intervals (CI) and variance of the KL divergence are estimated via bootstrapping \cite{carpenter_bootstrap_2000}. Significance for the relative similarity of two pairs of distributions, e.g., the radiologist and DL model density distributions before and after adaptation, is estimated by comparing the KL divergences between the two pairs using a two-sided two-sample t-test. Statements involving directional information, e.g. “slightly underestimates”, are evaluated based on a one-sided Wilcoxon signed-rank test. \\

The comparisons between the DL model and radiologist density distributions for the same dataset are summarized in the Table~\ref{tab:density_dist_comp}. Comparisons of the relative similarity of the two distributions before and after adaptation (matrix calibration, 500 images) are also provided. \\

\begin{table}[htbp]
    \centering
    \caption{A summary of the Breast Imaging Reporting and Data System (BI-RADS) breast density distributions of
the DL model and the original reporting radiologists. Similarity between the distributions is characterized by the Kullback-Liebler (KL) divergence and Pearson’s chi squared test. Comparisons of the similarity before and after adaptation are calculated by comparing the KL divergences using a two-sided t-test.}
    \begin{tabular}{lcccc}
        \vspace*{1mm}\\
        \toprule
         ~ & \shortstack{Radiologist\\Distribution} & \shortstack{DL Model\\Distribution} & KL Divergence & P-value \\
         \midrule
         FFDM & \shortstack{A: 9.3\%, B: 52.0\%,\\C: 34.6\%, D: 4.0\%} & \shortstack{A: 8.5\%, B: 52.2\%,\\C: 36.1\%, D: 3.2\%} & \shortstack{0.0015\\~[0.0011, 0.0018]} & $<$ 0.001 \\
         \midrule
         Site~1 (Before) & \shortstack{A: 8.9\%, B: 49.6\%,\\C: 35.9\%, D: 5.6\%} & \shortstack{A: 10.4\%, B: 57.8\%,\\C: 28.9\%, D: 3.0\%} & \shortstack{0.02\\~[0.00, 0.03]} & 0.01 \\
         \midrule
         Site~1 (After) & & \shortstack{A: 5.9\%, B: 53.7\%,\\C: 35.9\%, D: 4.4\%} & \shortstack{0.008\\~ [-0.005, 0.015]} & \shortstack{0.24\\(before vs. after: 0.13)} \\
         \midrule
         Site~2 (Before) & \shortstack{A: 15.3\%, B: 42.2\%,\\C: 30.2\%, D: 12.3\%} & \shortstack{A: 5.7\%, B: 48.8\%,\\C: 36.4\%, D: 9.4\%} & \shortstack{0.056\\~[0.041, 0.068]} & $<$ 0.001 \\
         \midrule
         Site~2 (After) & & \shortstack{A: 16.9\%, B: 43.3\%,\\C: 29.4\%, D: 10.4\%} & \shortstack{0.0026\\~
[0.0011, 0.0035]} & \shortstack{0.047\\(before vs. after: $<$ 0.001)} \\
         \bottomrule
    \end{tabular}
    \label{tab:density_dist_comp}
\end{table}

Statistical significance was also calculated for other comparisons related to the density distributions. For the Site~1 SM test set, the DL model slightly underestimates the breast density relative to the radiologists (P~$<$~.001). For
the Site~2 SM test set, the DL model did not underestimate the breast density (P~=~.99). The density distribution for the DL model for Site~2 is more similar to the radiologist distributions for Site~1 compared with the radiologist density distribution for Site~2 (Site~1 FFDM: KL~=~0.03 [95\%~CI:~0.02,~0.05], P~=~.03; Site~1 SM: KL~=~0.02 [95\%~CI:~-0.02,~0.03],~P~=~.01). This suggests that the model could have learned a prior density distribution from the Site~1 FFDM dataset that may not be optimal for Site~2 where patient demographics are different.

\section{Consistency of Image-level Predictions}\label{sec:appendix_consistency}

The exam-level predictions of the DL model are obtained by averaging the predicted probabilities for the four BIRADS breast density categories across all images in the exam. To better understand the consistency of the image-level predictions, the most probable BI-RADS breast density category for every image in an exam was considered for all exams in the FFDM test set. Typically, the breast density predictions for different views within an exam were consistent. For example, the predictions were the same for all four views 79.5\% (10546 of 13262) of the time. In almost all other cases, two BI-RADS breast density classes were predicted for the four views (20.5\%, 2715 of 13272). In all but one case, the two predicted classes were neighboring density classes (eg A and B). There was only one case where three distinct density predictions were made and no cases where four distinct predictions were made. In the case where three distinct density predictions were made, a prior surgery had removed most of the breast tissue from the right breast.

\newpage
\section{Supplementary Figures}

\begin{figure}[htbp]
    \centering
    \begin{subfigure}[b]{0.25\textwidth}
        \includegraphics[width=\textwidth]{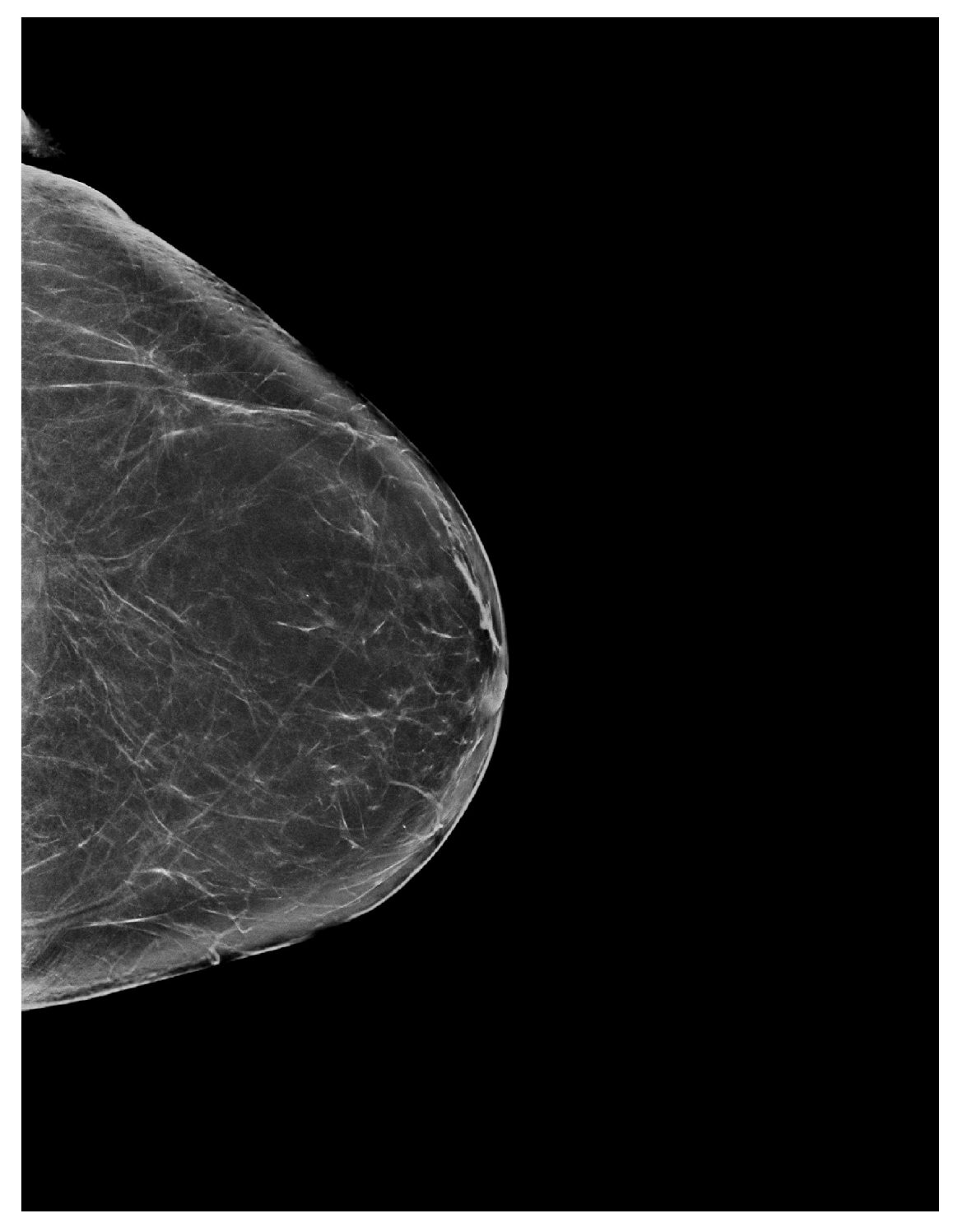}
        \caption{}
    \end{subfigure}%
    \begin{subfigure}[b]{0.25\textwidth}
        \includegraphics[width=\textwidth]{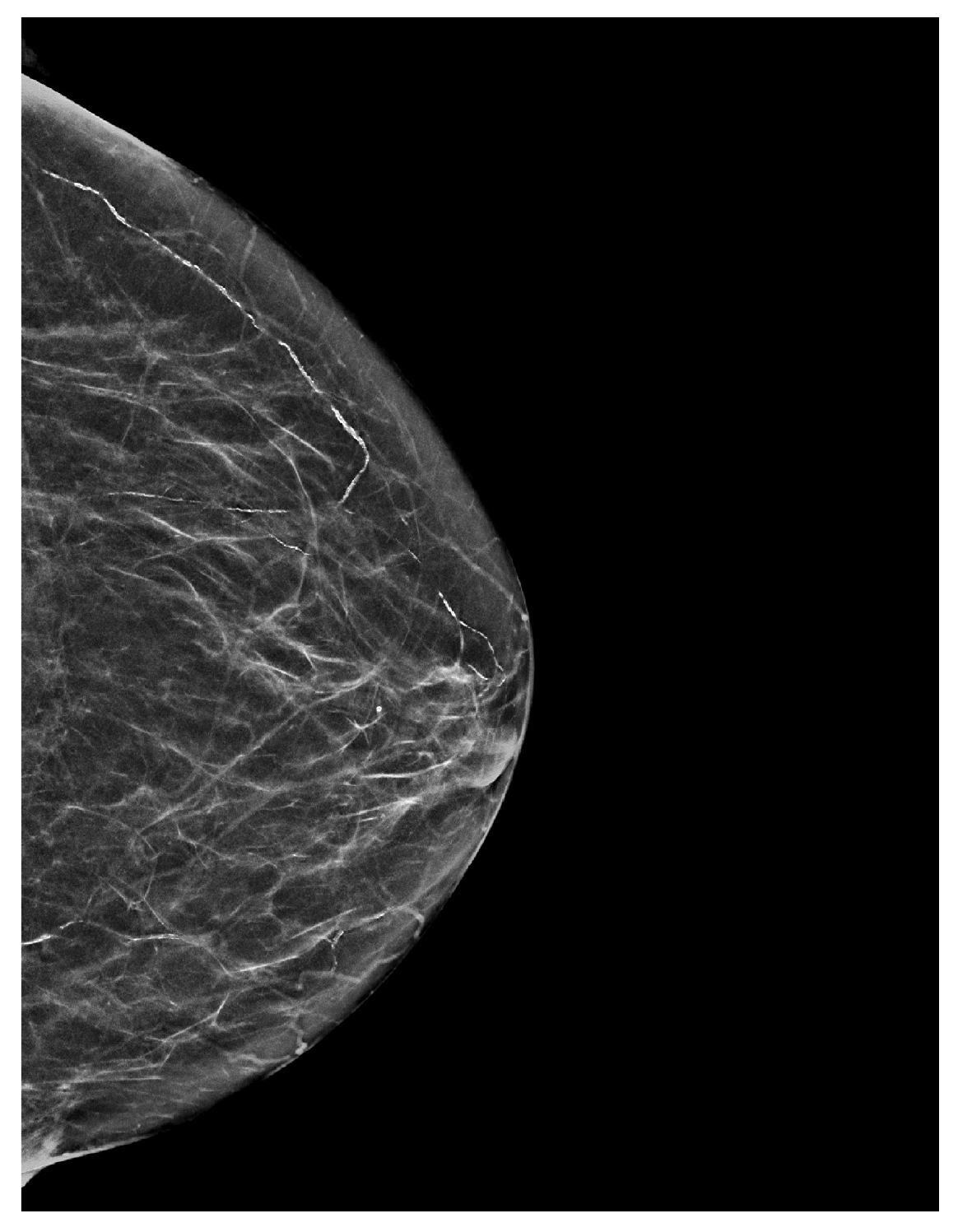}
        \caption{}
    \end{subfigure}%
    \begin{subfigure}[b]{0.25\textwidth}
        \includegraphics[width=\textwidth]{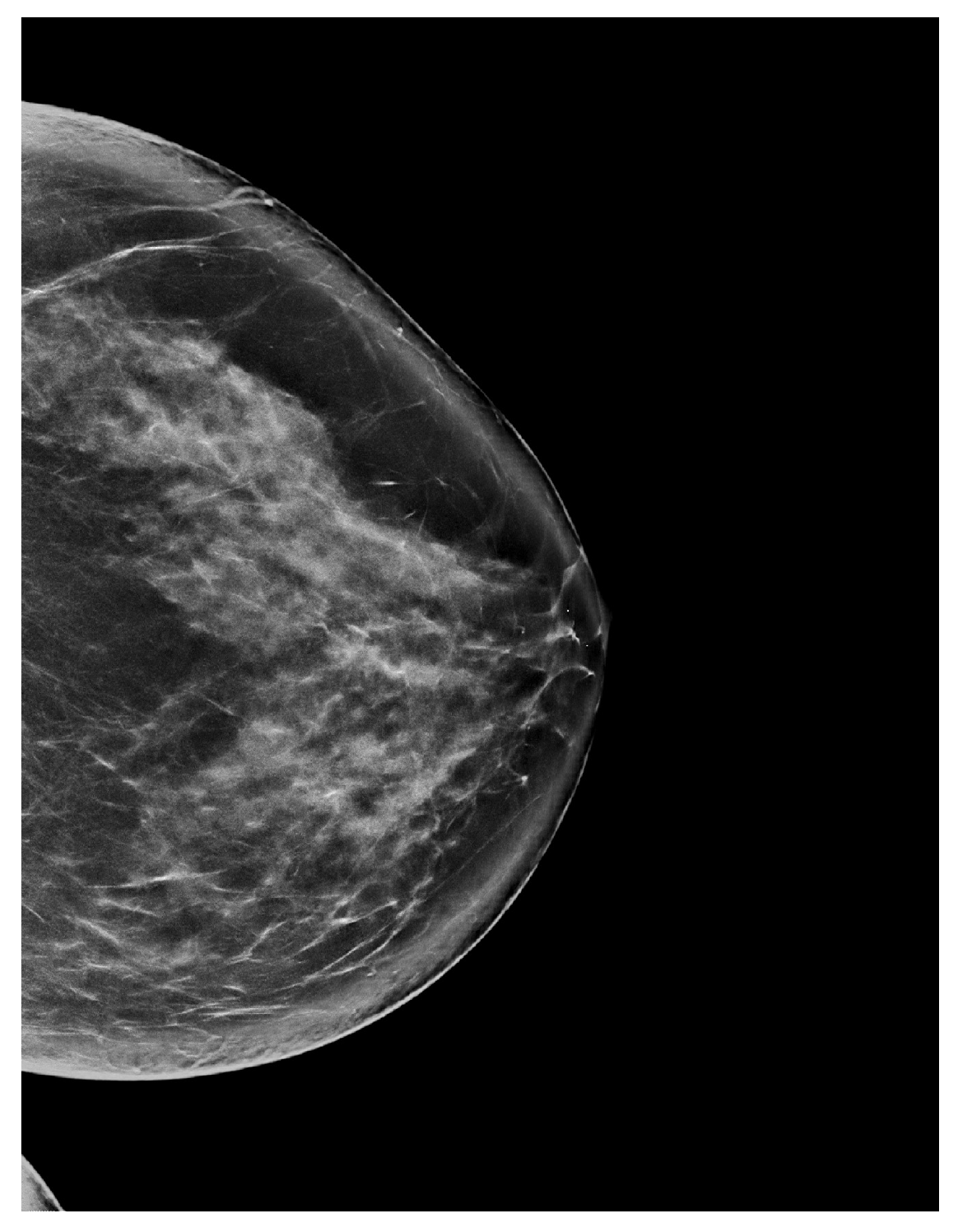}
        \caption{}
    \end{subfigure}%
    \begin{subfigure}[b]{0.25\textwidth}
        \includegraphics[width=\textwidth]{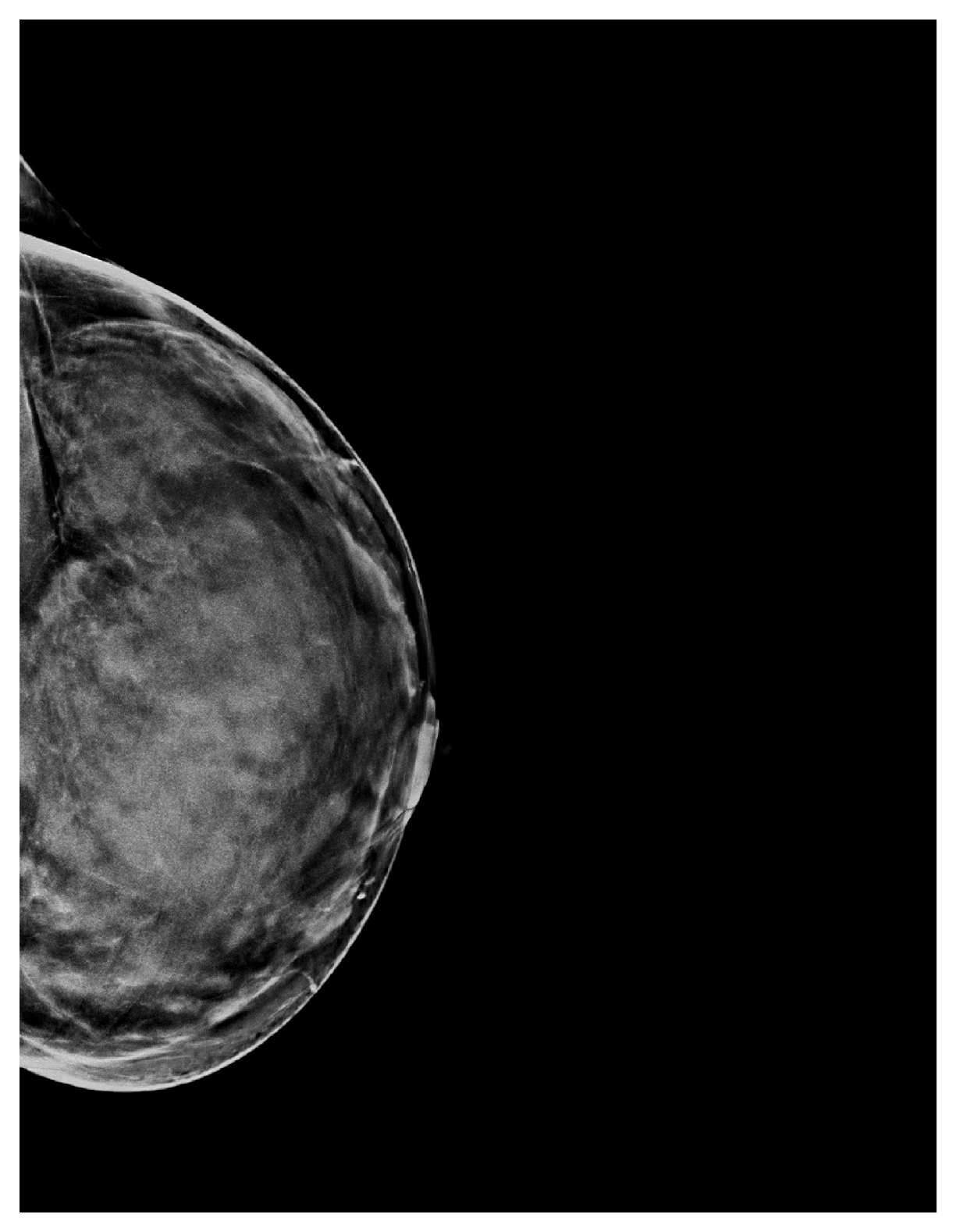}
        \caption{}
    \end{subfigure}%
    \caption{Example synthetic 2D mammography images derived from digital breast tomosynthesis exams for each of the four Breast Imaging Reporting and Data System (BI-RADS) breast density categories: A, BI-RADS A: almost entirely fatty, B, BI-RADS B: scattered areas of fibroglandular density, C, BI-RADS C: heterogeneously dense, and D, BI-RADS D: extremely dense. Images are normalized so that the grayscale intensity windows found in their Digital Imaging and Communications in Medicine (DICOM) headers range from 0.0 to 1.0.}
    \label{fig:example_density_cview}
\end{figure}

\begin{figure}[htbp]
    \centering
    \begin{subfigure}[b]{0.25\textwidth}
        \includegraphics[width=\textwidth]{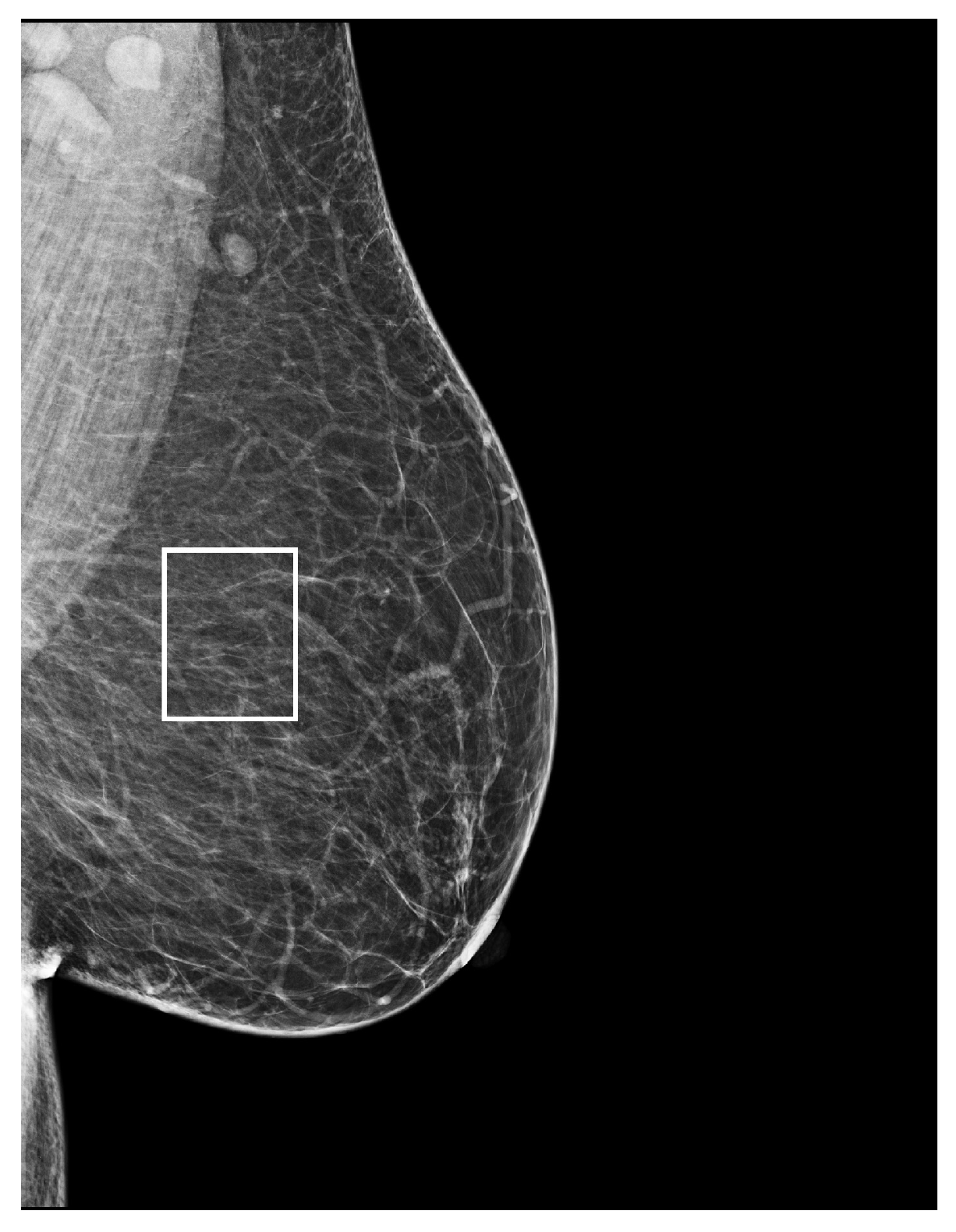}
        \subcaption{}
    \end{subfigure}%
    \begin{subfigure}[b]{0.25\textwidth}
        \includegraphics[width=\textwidth]{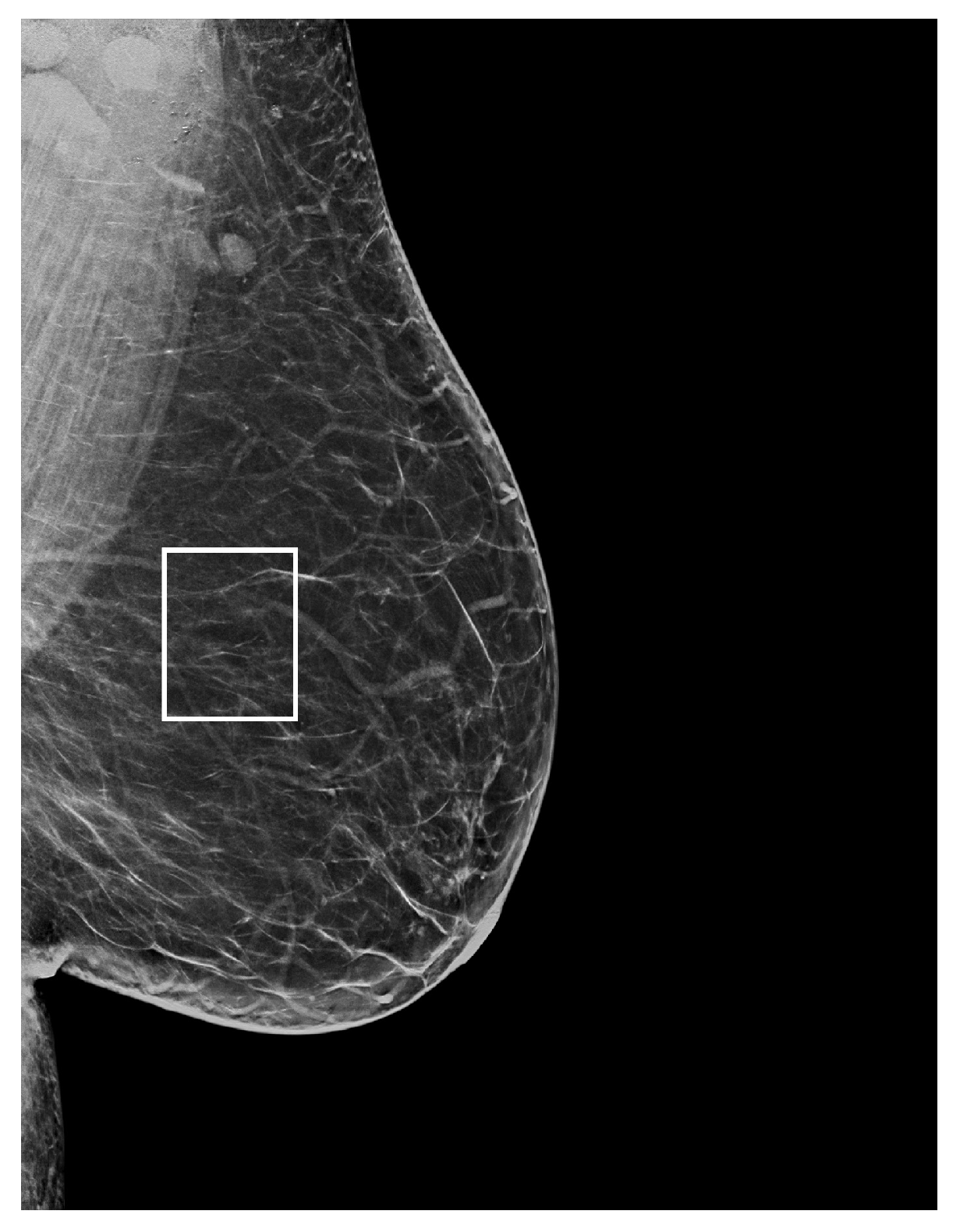}
        \subcaption{}
    \end{subfigure}%
    \begin{subfigure}[b]{0.25\textwidth}
        \includegraphics[width=\textwidth]{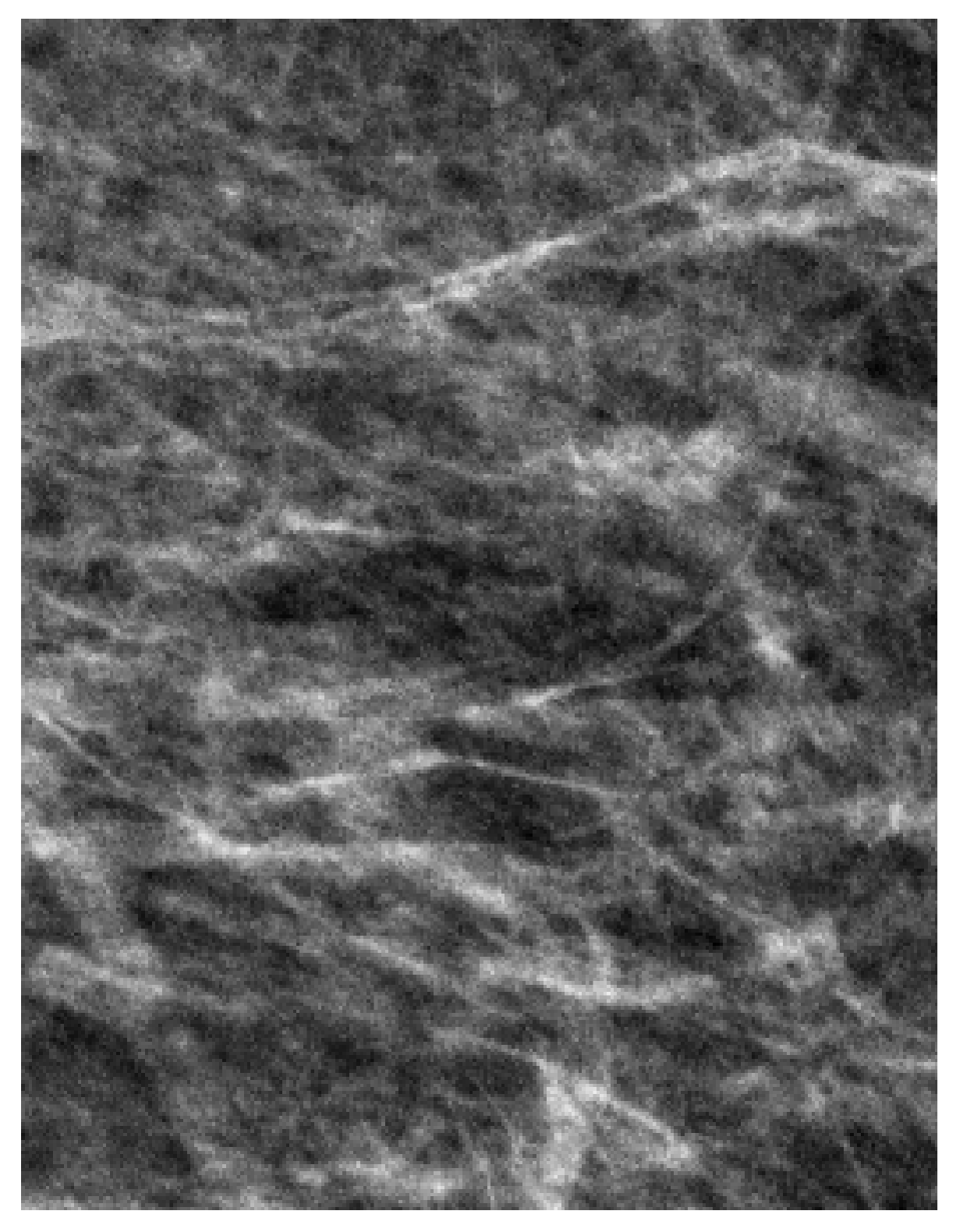}
        \subcaption{}
    \end{subfigure}%
    \begin{subfigure}[b]{0.25\textwidth}
        \includegraphics[width=\textwidth]{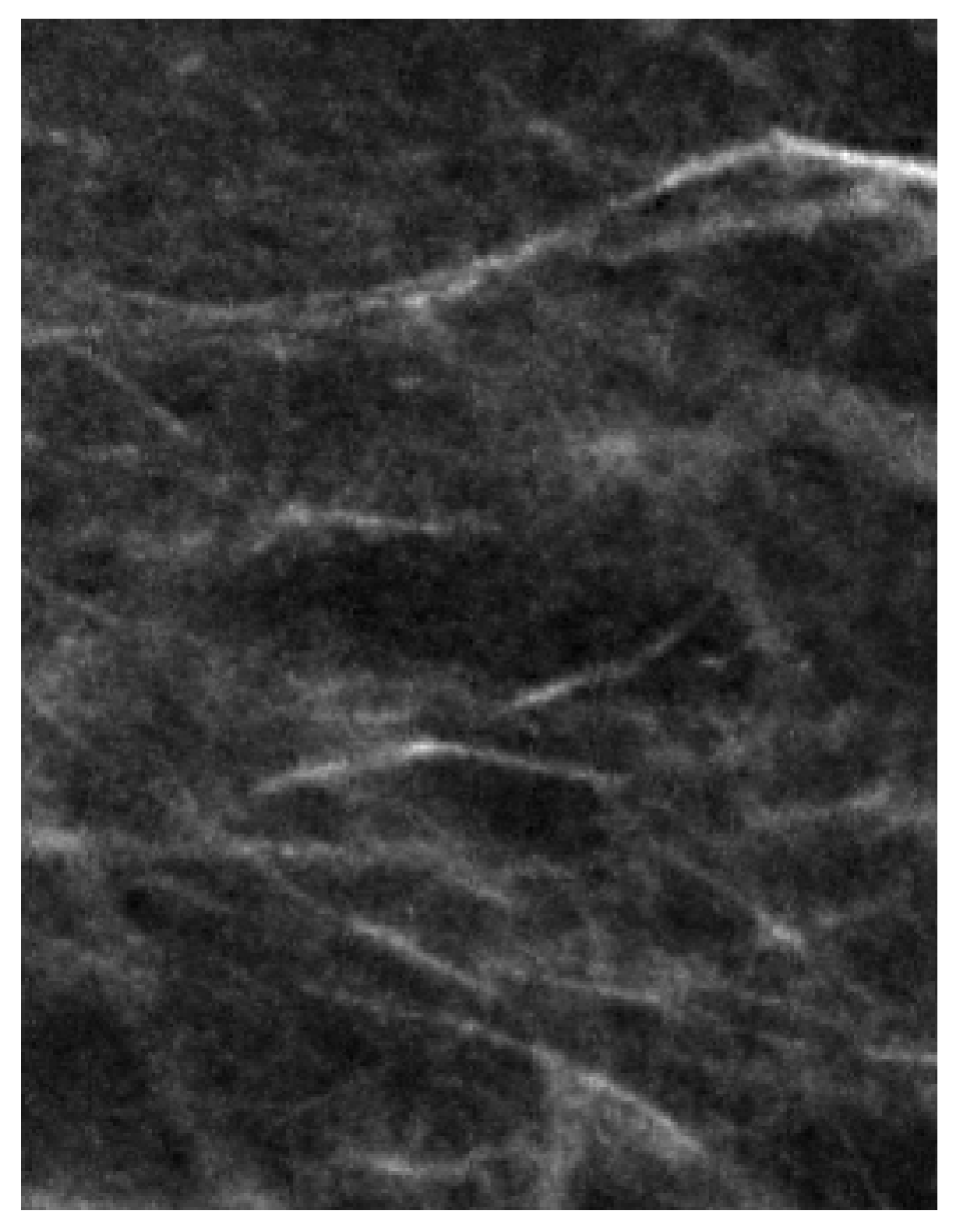}
        \subcaption{}
    \end{subfigure}%
    \caption{Comparison between A, a full-field digital mammography (FFDM) image and B, a synthetic 2D mammography (SM) image of the same breast under the same compression. A zoomed-in region, whose original location is denoted by the white box, is shown for both C, the FFDM image and D, the SM image to highlight the differences in texture and contrast that can occur between the two image types. Images are normalized so that the grayscale intensity windows found in their Digital Imaging and Communications in Medicine (DICOM) headers range from 0.0 to 1.0.}
    \label{fig:cview_ffdm_comparison}
\end{figure}

\FloatBarrier

\section*{Acknowledgements}
This work was supported in part by funding from Whiterabbit AI, Inc. Washington University has equity interests in Whiterabbit AI, Inc. and may receive royalty income and milestone payments from a “Collaboration and License Agreement” with Whiterabbit AI, Inc. to develop a technology evaluated in this research. In addition, the following authors are employed by and/or have equity interests in Whiterabbit AI, Inc.: T.P.M., S.S., B.M., J.S., M.P.S., S.P., A.L., R.M.H., N.G., D.S., and S.C.M. \\

The authors would like to thank Drs. Mark A. Anastasio, Catherine M. Appleton, and Curtis P. Langlotz for their insightful feedback and review of this manuscript. The authors would also like to thank Chip Schweiss for managing the research cluster with which this work was performed.


\bibliographystyle{abbrv}
\bibliography{density_adaptation}

\end{document}